\documentclass[colorlinks=true,linkcolor=blue]{article}
\usepackage{pgfplots}
\pgfplotsset{compat=1.18}

\usepackage{PRIMEarxiv}

\usepackage[utf8]{inputenc} % allow utf-8 input
\usepackage[T1]{fontenc}    % use 8-bit T1 fonts
\usepackage{hyperref}       % hyperlinks
\usepackage{url}    
\usepackage{comment}
\usepackage{booktabs}       % professional-quality tables
\usepackage{amsfonts}       % blackboard math symbols
\usepackage{nicefrac}       % compact symbols for 1/2, etc.
\usepackage{microtype}      % microtypography
\usepackage{lipsum}
\usepackage{fancyhdr}       % header
\usepackage{graphicx}       % graphics
\usepackage{amsmath}
\usepackage{threeparttable}
\usepackage{subcaption}
\usepackage{adjustbox}
\usepackage{bm}
\usepackage{algorithm}
\usepackage{algpseudocode}
\usepackage{hyperref}

\usepackage[section]{placeins}

\usepackage{amsmath, amssymb}      % simboli e ambiente matematico
\usepackage{tikz}                  % motore grafico
\usetikzlibrary{arrows.meta,
                positioning,
                calc,
                shapes.misc,
                shapes.geometric}   % librerie richiamate dal diagramma
\usepackage{sansmath}              % per scrivere formule in sans serif

\graphicspath{{media/}}     % organize your images and other figures under media/ folder

\title{End-to-End Large Portfolio Optimization for Variance Minimization with Neural Networks through Covariance Cleaning 
}

\author{
  Christian Bongiorno\\
  Université Paris-Saclay, CentraleSupélec,\\
  Mathématiques et Informatique pour la Complexité et les Systèmes, \\
  91190, Gif-sur-Yvette, France.\\
  \texttt{\href{christian.bongiorno@centralesupelec.fr}{christian.bongiorno@centralesupelec.fr}}
   \And
  Efstratios Manolakis \\
  Dipartimento di Fisica e Astronomia “Ettore Majorana”, \\
  Universitá di Catania,\\
  Via Santa Sofia, 64, 95123, Catania, Italy.\\
  \And
  Rosario Nunzio Mantegna\\
  Dipartimento di Fisica e Chimica,\\
  Università degli Studi di Palermo,\\
  Palermo, Italy. \\
  Complexity Science Hub,\\
  Metternichgasse 8\\
1030 Vienna, Austria. \\
}

\begin{document}
\maketitle

\begin{abstract}
We develop a rotation-invariant neural network that provides the global minimum-variance portfolio by jointly learning how to lag-transform historical returns and marginal volatilities and how to regularise the eigenvalues of large equity covariance matrices. This explicit mathematical mapping offers clear interpretability of each module’s role, so the model cannot be regarded as a pure black box. The architecture mirrors the analytical form of the global minimum-variance solution yet remains agnostic to dimension, so a single model can be calibrated on panels of a few hundred stocks and applied, without retraining, to one thousand US equities, a cross-sectional jump that indicates robust generalization capability. The loss function is the future short-term realized minimum variance and is optimized end-to-end on real returns. In out-of-sample tests from January 2000 to December 2024, the estimator delivers systematically lower realized volatility, smaller maximum drawdowns, and higher Sharpe ratios than the best competitors, including state-of-the-art non-linear shrinkage, and these advantages persist across both short and long evaluation horizons despite the model’s training focus is short-term. Furthermore, although the model is trained end-to-end to produce an unconstrained minimum-variance portfolio, we show that its learned covariance representation can be used in general optimizers under long-only constraints with virtually no loss in its performance advantage over competing estimators. These advantages persist when the strategy is executed under a highly realistic implementation framework that models market orders at the auctions, empirical slippage, exchange fees, and financing charges for leverage, and they remain stable during episodes of acute market stress.
\end{abstract}

%keywords can be removed

%Sections
\section{Introduction}

% What is portfolio management
Portfolio management has been fundamental to financial practice since the earliest securities markets emerged. Following Markowitz’s introduction of Modern Portfolio Theory (MPT) \cite{markowitz1952portofolio}, academic and industry efforts have prioritized the construction of portfolios that optimize risk‐adjusted returns. This framework employs historical return data and estimates of pairwise asset dependencies, typically captured by covariance matrices, to determine asset weightings. 

Classic portfolio optimization proceeds in two logical steps: parameter estimation followed by the optimization itself. The optimization phase depends on forecasts of asset return vectors and their covariance matrix. Predicting these quantities is notoriously difficult, and a vast literature seeks either to improve forecast accuracy or to develop stochastic frameworks that explicitly accommodate their uncertainty. Forecasting returns is especially challenging because it reflects the degree of market efficiency observed in financial markets \cite{fama1970efficient}. Estimating covariances likewise involves fundamental statistical estimation issues \cite{stein1975estimation} and must account for their time-varying nature. Nonetheless, covariance forecasting is relatively more tractable, since improvements in the covariance estimate cannot be directly translated into profit in simple trading strategies without resort to complex financial instruments.

The literature on machine learning–based portfolio optimization is vast and rapidly expanding \cite{lee2024overview,zhang2020deep}; a comprehensive review is provided in Sec.~\ref{sec: Related works}, and here we summarize only high-level themes.  Existing studies can be grouped into three main categories: span predict-then-optimize pipelines \cite{mesquita2020dynamic}, end-to-end differentiable frameworks \cite{butler2023integrating,costa2023distributionally,uysal2024end}, and Reinforcement-Learning (RL) agents  \cite{chaouki2020deep,jang2023deep,yan2024reinforcement}.  Overall, machine learning–based portfolio optimization is highly challenging with several limitations still to be overcome. In fact, some studies employ small or static asset universes that limit evidence of scalability and robustness \cite{mesquita2020dynamic,zhang2020deep,huang2024enhancing,uysal2024end}. Benchmarks are in some cases infrequent or sub-optimal, making it difficult to compare performance improvements  across studies \cite{lee2024overview}. Moreover, empirical protocols can be sensitive to  data leakage, for instance through survivorship-restricted investable universes \cite{huang2024enhancing,costa2023distributionally,yu2025optimization}  or, more generally, through look-ahead bias in feature construction and evaluation. These challenges constitute the principal obstacles to the practical deployment of machine learning in portfolio optimization.

%Contribution: Operationally 
Here we propose an end-to-end neural network for Global Minimum-Variance (GMV) portfolio optimization, designed to operate in an asset-selection setting that closely matches professional investment practice. The approach is dimensionally agnostic: a model calibrated on a few hundred stocks can be deployed on substantially larger universes without retraining. We evaluate performance using a realistic backtesting protocol that reproduces the operational life cycle of a cash-and-margin account, including long-only constraints, transaction costs, and other trading frictions. In Out-Of-Sample (OOS) tests on U.S. equities (2000--2024), the resulting portfolios achieve lower realized volatility, smaller drawdowns, and higher Sharpe ratios than widely used state-of-the-art covariance estimators, while scaling reliably from a few hundred to approximately 1{,}000 assets.

%Contribution: Methodologically 
Methodologically, we recast the classical GMV solution as a differentiable rotation-invariant pipeline and implement its key steps as a modular neural architecture. Returns for $n$ assets over $\Delta t_\textrm{in}$ observations are first processed by a learnable lag-transformation with soft clipping, after which a multivariate denoising block outputs a cleaned correlation structure while adapting to the aspect ratio $q_\textrm{in}:=n/\Delta t_\textrm{in}$, in parallel, a feed-forward module estimates inverse marginal volatilities. All components are trained jointly using the realized OOS variance of the resulting GMV portfolio as the learning signal, so that preprocessing, volatility scaling, and eigenvalue regularization are optimized end-to-end rather than selected a priori. The main architectural challenge lies in the multivariate denoising stage, where symmetry and validity constraints inherent to correlation matrices severely restrict admissible parameterizations. By embedding these constraints through a rotation-invariant design and exploiting Random Matrix Theory (RMT)-informed spectral regularities, the learned parameters can leverage known symmetries and dependencies, facilitating transfer across asset universes of different dimensionality. 
%Section~\ref{sec:NN} details the architecture, while Section~\ref{sec:Training} describes the training protocol and the empirical evaluation.

% Covariance cleaning and portfolio optimization
A key aspect of our model is associated with the determination of the covariance matrix.  When estimated from finite samples, empirical covariance matrices tend to be ill‐conditioned and sample covariance presents estimation errors \cite{stein1975estimation,laloux1999noise,ledoit2004well}, often resulting in suboptimal portfolio allocations. A widely adopted remedy is to apply Rotation Invariant Estimators (RIEs), which adjust the sample eigenvalues while retaining the sample eigenvectors \cite{bun2017cleaning}. Within this framework, optimality is measured by the Mean Squared Error (MSE), equivalent to the Frobenius norm distance between the estimator and the true covariance matrix. The theoretical optimum is the so‐called Oracle Estimator (OE), defined as the covariance matrix that minimizes this Frobenius distance. In practice, however, the OE cannot be realized, since it requires prior knowledge of the unknown true covariance.

One of the earliest covariance regularization techniques is eigenvalue clipping. Laloux et al.~\cite{laloux1999noise} showed that the bulk of the eigenvalue spectrum of empirical financial covariance matrices closely follows the Marchenko–Pastur distribution for random noise. Eigenvalue clipping therefore replaces all sample eigenvalues that lie within this noise bulk by their average value. Later, Ledoit and Wolf \cite{ledoit2004well} introduced the Linear Shrinkage (LS) estimator, which constructs a convex combination of the sample covariance matrix and a diagonal target matrix. This estimator remains well–defined when the number of assets $n$ exceeds the number of observations $\Delta t_\textrm{in}$. Its optimality in minimizing MSE holds strictly under the assumption that the true covariance matrix is drawn from an inverse–Wishart distribution \cite{lamrani2025optimal}.
In the last decade, Ledoit and collaborators introduced the NonLinear Shrinkage (NLS) estimator as an optimal method for covariance regularization \cite{ledoit2012nonlinear}. Deriving the analytical form of the Ledoit–P\'ech\'e (LP) estimator is challenging because it requires detailed knowledge of the sample eigenvector distribution \cite{ledoit2011eigenvectors}. The Quantized Eigenvalues Sampling Transform (QuEST) delivers the most accurate numerical implementation of the LP estimator by discretizing the limiting spectral distribution \cite{ledoit2015spectrum}, but it incurs substantial computational cost. To alleviate this burden, the direct kernel estimator \cite{ledoit2017direct} and the inverse–Wishart regularization approach \cite{bun2017cleaning} offer closed-form approximations that closely mimic the LP solution. More recently, Ledoit and Wolf proposed the Quadratic-Inverse Shrinkage (QIS) estimator \cite{ledoit2022quadratic}, which is considered the state-of-the-art technique in financial covariance filtering by combining computational efficiency with high estimation accuracy.

% Problem in covariance filtering, why the theory cannot be applied on the real world
Although the theoretical framework for covariance filtering is well established, its practical implementation remains challenging due to the pronounced non-stationarity of financial covariances. Market regimes evolve continuously under shifting economic conditions and investor behavior, so a covariance matrix estimated from past returns may poorly reflect current asset dependencies. Moreover, NLS theory assumes that sampling noise is the sole source of discrepancy between the sample and true covariance matrices and relies on the thermodynamic limit in which the number of assets $n$ and observations $\Delta t_\textrm{in}$ both tend to infinity with fixed aspect ratio $q$. In practice, extending $\Delta t_\textrm{in}$ to satisfy this limit only exacerbates non-stationarity, as longer historical windows encompass multiple structural shifts. Finally, the optimality of the LP estimator depends on the existence of finite twelfth-order moments of asset returns \cite{ledoit2011eigenvectors}, yet empirical evidence indicates that equity returns exhibit heavy tails with tail exponents near four, rendering higher-order moments effectively unbounded.
% CV approach to filter the covariance matrix
Methods based on $k$-fold Cross‐Validation (CV) have been proposed to relax the stringent assumptions of nonlinear shrinkage theory \cite{bartz2016advances,lam2016nonparametric,lam2018nonparametric,lamrani2025optimal,reigneron2020agnostic}. These CV estimators achieve accuracy comparable to more complex schemes while permitting straightforward adaptations to specific data features. For example, Tan et al.~\cite{tan2025estimation} introduced an exponential decay factor into the return series within the CV framework. By applying exponential weights, one can maintain a long calibration window, approaching the thermodynamic limit, while emphasizing the most recent observations. This modification partially resolves the bias-variance trade‐off caused by non‐stationarity and structural shifts in financial covariances.

% Hierarchical clustering
An alternative to RIEs employs hierarchical clustering to filter the empirical correlation matrix \cite{tumminello2010correlation,bongiorno2021covariance,bongiorno2022reactive}. Unlike RIEs, clustering‐based filters modify both eigenvalues and eigenvectors by aggregating assets into empirically determined groups. This approach exploits the fact that financial assets are often organized into a clear hierarchy of sectors and sub‐sectors \cite{mantegna1999hierarchical}. Although it lacks full theoretical support, clustering filters have, in some instances, yielded superior out‐of‐sample portfolio performance compared with RIEs \cite{tola2008cluster,pantaleo2011improved,bongiorno2021covariance,bongiorno2022reactive,garcia2023two}. By enforcing the observed hierarchical relationships, these methods produce the most parsimonious covariance approximation that exactly reproduces the data’s cluster structure.

% The DCC vs AO model, was a possible way to fix non-stationarity.
The Engle Dynamical Conditional Covariance (DCC) model \cite{engle2002dynamic} represents a principled attempt to accommodate non‐stationarity by modeling time‐varying correlations through Generalized AutoRegressive Conditional Heteroscedasticity (GARCH)‐type dynamics. It specifies separate equations for the conditional variances and then characterizes the evolving correlations of the standardized returns via a time-varying correlation matrix.
In principle, this approach should adapt to shifting market regimes and capture persistence in asset comovements. In practice, however, DCC is computationally intensive, often suffers from convergence issues, and yields OOS portfolios whose risk–return profiles rarely surpass those obtained with simpler static filters \cite{bongiorno2024covariance}. In contrast, the Average Oracle (AO) method \cite{bongiorno2023filtering,bongiorno2024covariance} abandons time‐varying adjustments altogether. It fixes a set of eigenvalues calibrated over an extended historical window and applies them uniformly to all subsequent covariance estimates. This static correction ignores short‐term noise but captures the enduring structure of the market. Empirical tests show that AO consistently outperforms both DCC‐based and other NLS techniques, suggesting that, under strong non‐stationarity, long‐term averages provide more reliable covariance estimates than complex dynamic models \cite{bongiorno2024covariance,fermanian2024model}.

% Neural Networks as a possible solution to the problem
The AO method effectively serves as a zero‐order baseline, applying fixed eigenvalue corrections without any attempt to anticipate future shifts. An ideal covariance estimator, by contrast, would move beyond denoising past returns and actively predict forthcoming covariance dynamics. NNs naturally suggest themselves for this role, since they can model highly nonlinear relationships between input data and desired outputs. 
Two main strategies can be used when training an NN for GMV optimization. The first is based on learning a covariance matrix estimator, which is then used to compute the portfolio weights with a Quadratic Programming (QP) minimization. This approach has the advantage of being versatile and adaptable to different minimization problems, especially if they are convex and optimally solvable. However, it has the drawback of requiring the estimation of a covariance matrix, which needs to be positive definite. In addition, the loss function should be taken as some type of distance between the predicted covariance matrix and the population covariance matrix, usually the Frobenius distance. This presents a challenge, as the Frobenius distance is not a good metric for covariance matrices in practical portfolio optimization problems \cite{bongiorno2025quantifying,bongiorno2023non}. The second approach is to train an NN to predict portfolio weights directly. This method might allow for a straightforward optimization of the desired loss function if the NN architecture is designed in a tailored way. Depending on the architecture, the NN can internally store a representation of the covariance matrix or some simplified version. In this paper, we adopt the latter approach.

% Our contribution
At the heart of our architecture lies the correlation‐denoising module (Sec.~\ref{subsec:EigenvalueCleaningNN}), which implements a RIE of the correlation matrix.  By applying a RIE, one forces all permutational symmetries already encoded in its mathematical formulation.  To perform this eigenvalue filtering efficiently, the symmetry of the eigenvalue transformation itself must be respected.  We therefore take inspiration from the well‐known mapping of sample eigenvalues to a one‐dimensional Coulomb gas \cite{forrester2010log}, in which each eigenvalue behaves as repelling identical charges confined by the population spectrum.  In our implementation, the ordered sequence of sample eigenvalues is processed through a bidirectional LSTM that enforces permutation equivariance and captures the local interactions among neighboring ``charges''.  The network outputs a $q$-dependent NLS function that adjusts each sample eigenvalue while leaving the empirical eigenvectors intact.  Two auxiliary modules prepare the raw return data for this denoising step.  A learnable lag‐transformation network weights look-back lags and applies soft clipping to control temporal influence and outliers (Sec.~\ref{subsec:ReturnsReweighting}).  A marginal‐volatility network converts each asset’s sample standard deviation into an inverse volatility scale (Sec.~\ref{subsec:DeepNN}).  These modules work in concert to present the LSTM with input representations whose temporal dependencies and scales are optimally calibrated.  Finally, the cleaned eigenvalues recombine with the inverse marginal volatilities to form an estimated inverse covariance matrix, from which GMV portfolio weights follow in closed form.  Calibrating all modules jointly to minimize realized OOS variance enables the model to predict future covariance dynamics under non‐stationary market regimes.  This end‐to‐end design thus overcomes the limits of Frobenius‐norm–based shrinkage and static filters and delivers a dynamically adaptive estimator of portfolio risk.

\subsection{Literature Review}\label{sec: Related works}
The literature on NN portfolio optimization spans three methodological strands: predict-then-optimise pipelines, end-to-end differentiable architectures, and RL agents. We review some representative studies from every strand, briefly highlighting some design choices.

In the predict-then-optimise strand, Mesquita et al.,\cite{mesquita2020dynamic} combine a Heterogeneous AutoRegressive filter with a multilayer perceptron to deliver positive-definite covariance forecasts. They apply their method to a portfolio of five equities without providing evidence of scalability. Reis et al.,\cite{reis2025deep} train a 3-D CNN–BiLSTM–attention model on a 14-asset mixed universe before retraining it on 260 exchange-traded funds, showing scalability of the method. However, the relative contribution of fixed-income ETFs to the unconstrained GMV allocations is not explicitly discussed. Huang et al.,\cite{huang2024enhancing} first screen equities with an AGC-CNN that fuses OHLC images, graph convolutions, and self-attention, then feeds the survivors to a GMV allocator, obtaining improved Sharpe and Sortino ratios on CRSP stocks (2012–2023). The study is performed on the universe of 128 survivor firms of the considered period.

End-to-end differentiable frameworks embed the optimiser within the network so that forecasting and allocation are learned jointly. Butler and Kwon,\cite{butler2023integrating} place the Markowitz module inside the computational graph via implicit differentiation of the KKT system, achieving lower OOS costs than those obtained with ordinary least squares. The model is trained on a 24-contract futures universe by tuning a single exponential moving-average covariance window and with a fixed risk-aversion coefficient. Zhang et al.,\cite{zhang2020deep} train an LSTM to map price histories of four index ETFs directly to long-only weights, maximising the Sharpe ratio. The method withstands the early-2020 Covid-19 shock. The set of assets comprises a relatively small ETF set.  Costa and Iyengar,\cite{costa2023distributionally} embed a distributionally robust mean–variance layer that back-propagates through investor-risk and ambiguity-set parameters, outperforming equal-weight and classical pipelines on 2000–2021 US equities. The sample spans 20 surviving S\&P 500 stocks. Uysal et al.,\cite{uysal2024end} integrate a differentiable risk-budgeting layer and raise the Sharpe ratio from 0.79 for nominal risk-parity to 1.24 for a portfolio of seven US ETFs (2011–2021).

RL-based studies treat portfolio choice as a sequential-decision task. Chaouki et al.,\cite{chaouki2020deep} use Deep Deterministic Policy Gradients on stylised markets with analytically known optima, showing agents that recover no-trade bands and threshold rules. Jang and Seong,\cite{jang2023deep} fuse technical indicators and long-horizon correlations through 3-D CNNs and tensor decomposition to guide an RL agent over 29 Dow components (2008–2019), achieving higher Sharpe ratios and lower drawdowns. The study is performed assuming zero slippage and fixed fees while excluding delisted firms. Yan et al.,\cite{yan2024reinforcement} unite CNN and LSTM encoders with deterministic policy gradients to beat mean-reversion and moving-average baselines on cryptocurrencies and Chinese A-shares. Yu et al.,\cite{yu2025optimization} embed a mean–variance layer inside an RL agent that reweights discrete-Fourier return components, outperforming equally weighted portfolio and static Markowitz on 47 CSI-300 survivors (2014–2024).

Across the differnt strands, three areas of improvement emerge: (i) asset universes should be increased to make applicability more realistic, dampening estimation risk, (ii) benchmark selection should be expanded and made less idiosyncratic to promote like-for-like comparisons, and (iii) careful planning of experimental protocols to avoide leakage of future information or understate implementation costs. We address these issues by training and testing on the full high- and mid-cap US equity universe, benchmarking against state-of-the-art minimum-variance strategies, and enforcing look-ahead-free splits with realistic frictions, so that observed performance gains reflect genuine model capability rather than experimental artefacts.

\section{Mathematical Background}
\subsection{GMV Portfolios under Finite-Sample Risk-Mode Uncertainty}\label{subsec:GMVP}
% Paragraph 1 – set-up and optimisation
The section motivates why finite-sample estimation error inflates OOS portfolio risk in high dimensions and interprets this effect in terms of noisy risk modes. For analytical tractability, we show a closed-form reference expression for the resulting risk inflation under multivariate normal and i.i.d.~returns. These assumptions are used only for the benchmark formula and are not required by our covariance-filtering method or by the empirical analysis, which is conducted on real equity returns that are non-Gaussian and exhibit heavy tails and conditional heteroskedasticity.

Consider an idealised setting in which an investor observes $\Delta t_{\mathrm{in}}>n$ independent realisations of the $n$-dimensional return vector $\boldsymbol{r}_{t} = \left\{ r_{t1}, r_{t2}, \ldots, r_{tn} \right\}^\top$. We refer to these observations $\Delta t_{\mathrm{in}}$ as the In-Sample (IS) data and assume each $\boldsymbol{r}_t$ is drawn from a multivariate normal distribution with zero mean and covariance matrix $\boldsymbol{\Sigma}$. 
The task of the GMV portfolio is to find weights $\boldsymbol{w} = \left\{ w_1, w_2, \ldots, w_n \right\}^\top$ that minimise portfolio variance subject to the full-investment constraint $\boldsymbol 1^{\top}\boldsymbol w=1$.  
Replacing the unknown covariance matrix with its Maximum-Likelihood Estimator (MLE) $\widehat{\boldsymbol{\Sigma}} = \boldsymbol{\Sigma}_{\text{MLE}}$ yields the typical application of the GMV portfolio
\begin{equation}
\boldsymbol w=\frac{\widehat{\boldsymbol\Sigma}^{-1}\boldsymbol 1}{\boldsymbol 1^{\top}\widehat{\boldsymbol\Sigma}^{-1}\boldsymbol 1}.
\label{eq:GMVQP}
\end{equation}
Performance is later assessed on an independent OOS period of length $\Delta t_{\mathrm{out}}$, sampled from the same distribution $\mathcal N(\boldsymbol 0,\boldsymbol\Sigma)$.

% Paragraph 2 – finite-sample inflation
Sampling error inflates the realised risk of the GMV portfolio $\sigma^2_\text{PTF} = \boldsymbol{w}^\top \boldsymbol{\Sigma}_\mathrm{out} \boldsymbol{w}$.  Under multivariate normality, the expected OOS variance is
\begin{equation}
    \mathbb E\!\left[\sigma_{\mathrm{PTF}}^{2}\right]
\approx \frac{1}{\boldsymbol 1^{\top}\boldsymbol\Sigma^{-1}\boldsymbol 1}
\Bigl(1+\frac{n}{\Delta t_{\mathrm{in}} -n}\Bigr)
=\sigma_\star^{2}\!\left(1+\frac{q_\mathrm{in}}{1-q_\mathrm{in}}\right),
\label{eq:expected_portfolio_variance}
\end{equation}

where $q_{\mathrm{in}}:=n/\Delta t_{\mathrm{in}}$ and $\sigma_\star^{2}=(\boldsymbol 1^{\top}\boldsymbol\Sigma^{-1}\boldsymbol 1)^{-1}$ denotes the population GMV portfolio variance, i.e., an idealized portfolio that is built and tested on the population covariance $\boldsymbol{\Sigma}$ \cite{frahm2009dominating}.  
Note that the inflation factor depends only on the estimation window, and it diverges when $\Delta t_{\mathrm{in}} \to n$. Interestingly, $\Delta t_{\mathrm{out}}$ affects only the dispersion of the variance estimator.

Eq.~\eqref{eq:expected_portfolio_variance} highlights that, even under an idealised stationary Gaussian setting, OOS risk inflation is governed by the aspect ratio $q_{\mathrm{in}}=n/\Delta t_{\mathrm{in}}$. In a stationary environment, the standard remedy is to reduce $q_{\mathrm{in}}$ by increasing the estimation window $\Delta t_{\mathrm{in}}$, thereby letting sampling noise average out. For financial covariances, however, this prescription is intrinsically limited because the dependence structure evolves over time: enlarging $\Delta t_{\mathrm{in}}$ mixes heterogeneous market regimes and yields an estimator that is increasingly biased with respect to the covariance relevant for the subsequent holding period. In other words, non-stationarity limit the effective look-back horizon, because observations that are too old become statistically outdated for forecasting near-term risk.

This creates a structural tension in large-scale portfolio construction. Increasing $n$ is attractive because it enlarges the diversification set and may lower the population GMV variance $\sigma_\star^2$, yet non-stationarity constrains $\Delta t_{\mathrm{in}}$ to remain short, so $q_{\mathrm{in}}:=n/\Delta t_{\mathrm{in}}$ grows and Eq.~\eqref{eq:expected_portfolio_variance} predicts stronger finite-sample risk inflation. In this regime, the usual stabilisation strategy of extending $\Delta t_{\mathrm{in}}$ is problematic: it reduces sampling variance but increases bias by mixing heterogeneous regimes, whereas shortening $\Delta t_{\mathrm{in}}$ limits this bias but makes $\widehat{\boldsymbol{\Sigma}}$ ill-conditioned and $\widehat{\boldsymbol{\Sigma}}^{-1}$ unstable. Our regularisation should therefore be read as a pragmatic response to this constraint: it does not forecast regime shifts, but rather emphasises the most recent information and aims to extract a more reliable covariance signal from limited, potentially more relevant samples.

% Paragraph 3 – motivation for a spectral view
Since most covariance-cleaning procedures act through spectral modifications of the MLE correlation or covariance matrices, it is instructive to express the GMV portfolio in the eigenbasis and examine the contribution of each risk mode.

% Paragraph 4 – covariance eigen-basis
Let $\boldsymbol\Sigma=\boldsymbol U\boldsymbol Z\boldsymbol U^{\top}$ with eigenvalues $\zeta_k$ and eigenvectors $\boldsymbol u_k$.  The GMV portfolio can be decomposed as
\begin{equation}\label{eq:GMVQP_eigen}
\boldsymbol w\propto\sum_{k=1}^{n}\frac{c_k}{\zeta_k}\,\boldsymbol u_k ,
\qquad
c_k:=\boldsymbol u_k^{\top}\boldsymbol 1 =  \sum_{i=1}^n u_{ik}.
\end{equation}
We orient each eigenvector so that 
$c_k\ge 0$; changing the sign of an eigenvector leaves $\boldsymbol\Sigma$ unchanged, so this choice is harmless.  
The coefficient $c_k$ gauges how closely the eigen-portfolio $\boldsymbol u_k$ points toward the budget line $\boldsymbol 1$.   If $c_k=0$ the mode is orthogonal to $\boldsymbol 1$ and has equal long and short exposure, hence no net investment.   Allocating wealth to such a direction therefore increases variance without helping to deploy funds, so the GMV portfolio gives preference to modes with $c_k>0$.   Eq.~\eqref{eq:GMVQP_eigen} makes this explicit: eigen-portfolio $\boldsymbol u_k$ enters with weight $c_k/\zeta_k$, the product of its alignment with the budget and the inverse of its own variance $\zeta_k=\boldsymbol u_k^{\top}\boldsymbol\Sigma\boldsymbol u_k$.   Combining modes in this way spreads risk across orthogonal directions and achieves a total variance lower than that of any single eigen-portfolio.

% Paragraph 5 – correlation eigen-basis
Similarly, the eigendecomposition $\boldsymbol C=\boldsymbol V\boldsymbol\Lambda\boldsymbol V^{\top}$ of the correlation matrix leads to
\begin{equation}
w_i\propto\sum_{k=1}^{n}\frac{c_k}{\lambda_k}\frac{v_{ik}}{\sigma_i},
\qquad
c_k:=\sum_{i=1}^{n}\frac{v_{ik}}{\sigma_i},  \qquad i=1,\ldots,n;
\end{equation}
where $\sigma_i$ is the marginal volatility of the asset $i$.  
The scaling by $\sigma_i$ magnifies the influence of low-volatility assets and attenuates that of volatile ones.  Although the resulting eigen-portfolios are not more orthonormal, they are still orthogonal; therefore, the interpretation of $c_k$ and $\lambda_k^{-1}$ remains unchanged.

In practice, one does not observe the true eigenbasis but only its sample estimate from finite data.  RMT results indicate that when a population eigenvalue $\lambda_k$ is well separated from the rest of the spectrum, the corresponding sample eigenvector $\widehat{\boldsymbol v}_k$ tends to have a high overlap with the true $\boldsymbol v_k$ \cite{muirhead2009aspects}.  Unfortunately, such spectral isolation occurs primarily for the largest eigenvalues, which enter the global minimum‐variance portfolio with weights proportional to $1/\lambda_k$ and thus contribute little to risk reduction.  The smaller eigenvalues are tightly clustered in the bulk of the spectrum, causing their sample eigenvectors to mix and lose any precise orientation.  As a result, estimated eigen‐portfolios associated with these directions are dominated by sampling noise rather than true risk modes.

% Paragraph 6 – empirical shortcomings and their causes
Furthermore, empirically realised OOS portfolio variances often exceed the theoretical benchmark by an order of magnitude \cite{demiguel2009optimal,pantaleo2011improved}.  
Two features of equity data might explain most of this gap.  
First, returns are heteroskedastic and exhibit time-varying correlations, so that $\boldsymbol\Sigma_{\mathrm{in}}$ may differ substantially from $\boldsymbol\Sigma_{\mathrm{out}}$ \cite{engle1982autoregressive,Zumbach2003,best1991sensitivity}.  
Second, return distributions are heavy-tailed, well-approximated by a Student-t law with four to six degrees of freedom \cite{platen2008empirical,markowitz1996likelihood}.  
Because finite fourth moments are absent or unstable, the classical Wishart assumptions underlying RMT break down, amplifying estimation error \cite{bongiorno2023optimal}.

If short selling is forbidden, one adds the element-wise constraint $w_{i}\ge 0$ to the GMV optimization. The problem remains convex but no longer has a closed-form solution; however, it can still be optimally solved by numerical QP. Empirically, prior studies have reported that, under long-only constraints, portfolios constructed using the sample covariance matrix (i.e., the MLE) often deliver performance comparable to that obtained with covariance cleaning procedures; see Refs.~\cite{pantaleo2011improved,jagannathan2003risk}.

\subsection{Eigenvalue Cleaning for MSE Minimization}\label{subsec:EigenvalueCleaning}
Having established the problem and its connection to the portfolio’s eigen-risk modes, we now outline how conventional covariance-cleaning methods operate, thereby laying the groundwork for our NN–based cleaning procedure.

Let us first consider the case when the covariance estimator is taken as $\widehat{\boldsymbol{C}}=\boldsymbol{C}_\text{MLE}$.  Here  $\boldsymbol{C}_\text{MLE}$ is the MLE of the correlation matrix obtained from $\Delta t_\text{in}$ observations drawn from a multivariate distribution with covariance matrix $\boldsymbol{\Sigma} = \boldsymbol{D} \boldsymbol{C} \boldsymbol{D}$, with $\boldsymbol{D}=\text{Diag}(\sigma_1, \ldots, \sigma_n)$ being the diagonal matrix of the population assets' volatilities and $\boldsymbol{C}$ the population correlation matrix, and let $\boldsymbol{C}_{\text{MLE}}= \widehat{\boldsymbol{V}} \widehat{ \boldsymbol{\Lambda}} \widehat{\boldsymbol{V}}^{\top}$ be the spectral decomposition of the MLE. In what follows, we adopt the convention that a hat indicates quantities estimated via the MLE from sample data if not otherwise specified.
A RIE replaces the noisy eigenvalues with a deterministic function $\boldsymbol{f}$ while keeping $\widehat{\boldsymbol{V}}$ unchanged:
\begin{equation}
    \mathbf{\Xi}_{\boldsymbol{f}}
   := \widehat{\mathbf{V}}\,\boldsymbol{f}( \boldsymbol{C}_{\text{MLE}}\mid q)\,\widehat{\mathbf{V}}^{\top}.
\end{equation}
Assuming rotational invariance of the sampling process, the function $\boldsymbol{f}$ can be expressed in terms of the eigenvalues of the empirical correlation matrix
\begin{equation}\label{eq:frie}
   \boldsymbol{f}(\boldsymbol{C}_{\text{MLE}}\mid q) \mapsto \mathrm{Diag}(\boldsymbol{f}( \widehat{\boldsymbol{\lambda}}\mid q)).
\end{equation}
Because the map depends only on the empirical spectrum (and $q$), it is independent of the empirical eigenvectors and therefore basis-independent: any orthogonal similarity transform $\mathbf{O}\boldsymbol{C}_{\mathrm{MLE}}\mathbf{O}^{\top}$ yields the same eigenvalue correction, with the cleaned matrix rotating accordingly.

Notably, $\boldsymbol{f}$ is a vector-valued function, i.e., $ \boldsymbol{f}:\;\mathbb R^{n+1}\;\rightarrow\;\mathbb R^n$, such that
\begin{multline}\label{eq:vectvalue}
\boldsymbol f(\widehat{\lambda}_1,\dots,\widehat{\lambda}_n \mid q)
=
\left\{ f_1(\widehat{\lambda}_1,\dots,\widehat{\lambda}_n \mid q),\;\dots,\;f_n(\widehat{\lambda}_1,\dots,\widehat{\lambda}_n \mid q)\right\}, \\ \text{with} \quad f_k:\mathbb{R}^{n+1}\to\mathbb{R};
\end{multline}
which means that it is generally a non-separable map since each component may depend on the entire argument. Furthermore, $\boldsymbol{f}$ is permuational-equivariant; therefore, called $\tau \in \mathcal{S}_n$ a permutation operator, then 
\begin{equation} \label{eq:permeq}
    \tau\left( \boldsymbol{f}(\widehat{\boldsymbol{\lambda}}|q)\right)=\boldsymbol{f}\left(\tau(\widehat{\boldsymbol{\lambda}})|q\right), \quad \forall \tau \in \mathcal{S}_n.
\end{equation}
Here $\tau$ denotes a relabelling of eigenpairs. Permuting the entries of $\widehat{\boldsymbol{\lambda}}$ is implicitly accompanied by the same permutation of the associated eigenvectors (i.e., a column permutation of $\widehat{\mathbf{V}}$), so that the eigendecomposition of $\widehat{\mathbf{C}}$ is left unchanged. We note that Eq.~\eqref{eq:permeq} does not introduce any additional probabilistic assumption beyond the standard RIE framework. The sample eigendecomposition is defined only up to a relabelling of eigenpairs; therefore, any estimator that acts on the empirical spectrum and keeps the empirical eigenvectors must be consistent under such relabellings, which is exactly the permutation-quivariance property stated in Eq.~\eqref{eq:permeq}. These two structural properties of $\boldsymbol{f}$ (Eqs.~\eqref{eq:vectvalue}–\eqref{eq:permeq}) anticipate the constraints we impose on the NN architecture in Sec.~\ref{subsec:EigenvalueCleaningNN}.

The optimal $\boldsymbol{f}^\star(\widehat{\boldsymbol\lambda}\mid q)$ is defined as the minimiser of the Frobenius‐norm loss against the (unknown) population correlation matrix,
\begin{equation}\label{eq:eq:oracle_def}
    \boldsymbol f^\star(\widehat{\boldsymbol\lambda}\mid q)
\;:=\;\arg\min_{\boldsymbol{\lambda}\in\mathbb R^n}
\Bigl\|\boldsymbol C-\widehat{\boldsymbol{V}}\operatorname{Diag}(\boldsymbol{\lambda})\widehat{\boldsymbol{V}}^\top\Bigr\|_F^2.
\end{equation}
Because it requires knowledge of the (unknown) population correlation matrix $\boldsymbol{C}$, the exact solution of Eq.~\eqref{eq:eq:oracle_def} is called an oracle $\boldsymbol{f}^\mathrm{OE}(\widehat{\boldsymbol{C}},\boldsymbol{C})$ and is not implementable in practice.

A straightforward minimisation of Eq.~\eqref{eq:eq:oracle_def} yields the closed-form solution
\begin{equation}\label{eq:oracleclosed}
f^{\mathrm{OE}}_k(\widehat{\boldsymbol{C}},\boldsymbol{C})
   =\sum_{l=1}^{n}\lambda_{l}\,\psi_{kl},
\qquad
\psi_{kl}:=
          \sum_{i=1}^{n}(\widehat{v}_{ik}v_{il})^{2}.
\end{equation}
where $\psi_{kl}$ is the overlap between the $k$-th eigenvector of the empirical correlation matrix and the $l$-th eigenvector of the population correlation matrix.
Given that $\boldsymbol{\Psi}$ is an overlap between two reference frames, it is trivial to prove that it is a bi-stochastic matrix, i.e., the sum of each row and each column is equal to one. From this property comes that the oracle eigenvalues can be viewed as a weighted average of the population eigenvalues, where the weights are given by the overlap between the empirical and population eigenvectors.  For bulk indices $k$, the weights $\psi_{kl}$ are widely dispersed over the $l$ elements in the bulk, so that $f_k^\mathrm{OE}$ becomes an almost uniform average of many population eigenvalues within the bulk.  In a rough approximation, eigenvalue cleaning replaces the noisy fine structure of the bulk by a single effective level, while leaving the well-isolated high eigenvalues nearly untouched. 

Clearly, this estimator is not available in practice because $\boldsymbol{C}$ is unknown.
However, under rotational invariance, the expectation of both the population eigenvalues $\boldsymbol{\Lambda}$  and the overlap $\boldsymbol{\Psi}$ must depend only on the eigenvalues of the empirical covariance matrix $\widehat{\boldsymbol{\Lambda}}$ and the aspect ratio $q:=n/\Delta t_\textrm{in}$. Such expectation cannot be derived analytically in a closed form, but NLS theory provides an analytical approximation of $\boldsymbol{f}^\star \to\boldsymbol{f}^\mathrm{OE}
$ in the limit $n,\Delta t_\textrm{in}\to\infty$ with $q$ fixed \cite{ledoit2012nonlinear}.

It is critical to notice that the loss function (Eq.~\eqref{eq:eq:oracle_def}) used in the NLS theory is the Frobenius distance between the population correlation matrix and the empirical correlation matrix. This is not the same as the loss function used in a GMV portfolio optimization, which should be the realized OOS variance of the portfolio weights $\sigma^2_\text{PTF}$. Although in Ref.~\cite{ledoit2017nonlinear} the authors claim that both loss functions are equivalent, this is not true in practice. In Ref.~\cite{bongiorno2023non}, it was shown that the Frobenius-based OE is, in fact, not optimal for portfolio optimization.

\section{Neural Networks for Portfolio Optimization}\label{sec:NN}
% description of the NN architecture
The NN architecture that we propose is organized to replicate the analytical workflow of the GMV estimator. Embedding the algebraic symmetries of the GMV solution into the network topology both limits the number of free parameters and promotes robust OOS performance in large universes.

\begin{figure}
\centering
\includegraphics[width=1.0\textwidth]{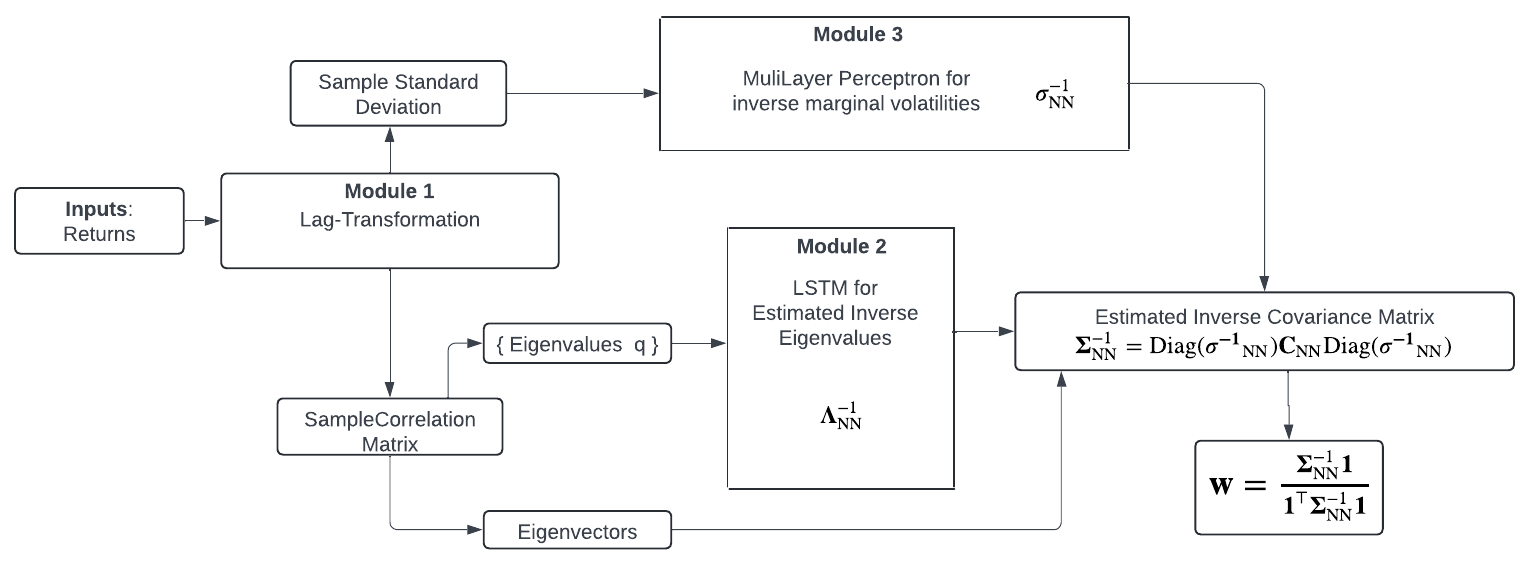}
\caption{Schematic representation of the proposed NN architecture.}\label{fig:NN_architecture}
\end{figure}

Fig.~\ref{fig:NN_architecture} highlights the three learnable modules of our architecture. First, the lag-transformation module (Model 1, Sec.~\ref{subsec:ReturnsReweighting}), applies a learnable temporal weighting kernel and a soft-clip to the raw return matrix $\boldsymbol{R}$, yielding the transformed returns $\widetilde{\boldsymbol{R}}$. Different lags can receive unequal weights; recent observations may be stressed, older observations may be down-weighted, or the reverse, depending on what minimizes the loss. The resulting sequence bifurcates into two parallel branches.

The lower branch standardizes $\widetilde{\boldsymbol{R}}$ to $\widetilde{\boldsymbol{Z}}$ and passes it to the eigenvalue-cleaning module (Model 2, Sec.~\ref{subsec:EigenvalueCleaningNN}). This block learns a non-linear shrinkage of the sample correlation matrix $\widetilde{\boldsymbol{C}}=\widetilde{\boldsymbol{Z}}\widetilde{\boldsymbol{Z}}^\top/\Delta t_\mathrm{in}$, producing a cleaned inverse eigenvalue matrix $\boldsymbol{\Lambda}_\mathrm{NN}^{-1}$.

The upper branch processes the vector of marginal volatilities $\left\{\widetilde{\sigma}_1,\ldots,\widetilde{\sigma}_n\right\}$ of $\widetilde{\boldsymbol{R}}$ through the marginal-volatility network (Model 3, Sec.~\ref{subsec:DeepNN}). This network transforms each standard deviation into its optimal inverse, generating the diagonal matrix $\boldsymbol{D}_\mathrm{NN}^{-1}= \operatorname{Diag}(\boldsymbol{\sigma}_\mathrm{NN}^{-1})$.

The two branches recombine to form the estimator of the inverse covariance matrix,
\begin{equation}
\boldsymbol{\Sigma}_\mathrm{NN}^{-1}= \boldsymbol{D}_\mathrm{NN}^{-1}\boldsymbol{V}_\mathrm{NN}
\boldsymbol{\Lambda}_\mathrm{NN}^{-1} \boldsymbol{V}_\mathrm{NN}^\top\boldsymbol{D}_\mathrm{NN}^{-1}.
\end{equation}
The matrix \(\boldsymbol{V}_{\mathrm{NN}}\) is obtained from the raw eigenvectors \(\widetilde{\boldsymbol{V}}\) of the lag‐transformed correlation \(\widetilde{\boldsymbol{C}}\) and projected so that the estimated correlation has ones on its diagonal.  In closed form,
\begin{equation}
    \boldsymbol{V}_\mathrm{NN} = \operatorname{Diag}\left(\operatorname{diag}\left( \widetilde{\boldsymbol{V}} \boldsymbol{\Lambda}_\mathrm{NN} \widetilde{\boldsymbol{V}}^\top\right)\right)^{-\frac{1}{2}} \widetilde{\boldsymbol{V}},
\end{equation}
where \(\mathrm{diag}(\cdot)\) extracts the diagonal vector and \(\mathrm{Diag}(\cdot)\) forms a diagonal matrix from that vector.  Finally, substituting $\boldsymbol{\Sigma}_\mathrm{NN}^{-1}$ into $\widehat{\boldsymbol{\Sigma}}^{-1}$ of Eq.~\eqref{eq:GMVQP} yields the portfolio weights.

The architecture is trained end-to-end by minimizing the realized OOS variance of the resulting portfolio. The loss function is
\begin{equation}\label{eq:loss}
\mathcal{L}(\boldsymbol{w}, \boldsymbol{\Sigma}_\mathrm{out})=
n \, \boldsymbol{w}^\top\boldsymbol{\Sigma}_\mathrm{out}\boldsymbol{w},
\end{equation}
where $\boldsymbol{\Sigma}_\mathrm{out}$ denotes the OOS realized covariance matrix of portfolio returns. The factor $n$ rescales the objective so that its magnitude is less sensitive to the cross-sectional dimension.

A reference implementation is available online and can be installed via PyPI \cite{rienet_code}.

\subsection{Lag‐Transformation Module}\label{subsec:ReturnsReweighting}
It is reasonable to assume that past observations do not influence the future risk uniformly. To make our NN lag aware, we design a specific module that, before the input returns reach the subsequent blocks, processes them through a learnable lag-transformation system that assigns a distinct transformation to every look-back lag. Classical risk models define a fixed exponential kernel \cite{zumbach2007riskmetrics,tan2025estimation}; our formulation generalizes that kernel by letting the NN determine its shape by training on the real data. 

Each raw return $r_{t i}$ is mapped to
\begin{equation}\label{eq:lagw}
\widetilde r_{t i}=\frac{\alpha_t}{\beta_t}
\tanh\bigl(252\,\beta_t r_{t i}\bigr),
\end{equation}
with $\boldsymbol{\alpha}=\{\alpha_{t}\}_{t=1}^{\Delta t_{\text{in}}}$ and $\boldsymbol{\beta}=\{\beta_{t}\}_{t=1}^{\Delta t_{\text{in}}}$ learned end-to-end together with the rest of the pipeline by back-propagation.  The specific functional form in Eq.~\eqref{eq:lagw} is primarily motivated by monotonicity and smooth differentiability, which supports stable back-propagation, by boundedness, which yields robustness to outliers through saturation, and by interpretability, since $\alpha_t$ controls the local linear scale while $\alpha_t/\beta_t$ sets the saturation level. This choice is not unique, and alternative monotone, differentiable, bounded parametrizations could be used without changing the overall framework.. Moreover, we stress that Eq.~\eqref{eq:lagw} defines an element-wise differentiable parametric transform, not a feed-forward Multi-Layer Perceptron (MLP). The coefficient $\alpha_{t}$ rescales observations at lag $t$.  The coefficient $\beta_{t}$ modulates their outlier sensitivity by controlling the saturation of the hyperbolic tangent.  When $\beta_{t}$ is small, the mapping is nearly linear ($\tanh(x)\approx x$ if $x$ is small) and the return enters almost unchanged; when $\beta_{t}$ is large, the mapping approaches a symmetric step that clips extreme values at $\pm\alpha_t/\beta_t$.  In this way, the module can emphasise or attenuate individual lags and dampen outliers simultaneously.

The factor $252$ annualises the input and stabilises gradient computations under single-precision arithmetic.  Because $\alpha_{t}$ and $\beta_{t}$ are updated jointly with all other network parameters, the module can learn any monotone or non-monotone lag profile, with the classical exponential decay arising as a special case.  The output transformed matrix $\widetilde{\boldsymbol{R}}$ serves as the input to all subsequent modules.

\subsection{Eigenvalue Cleaning with Neural Networks}\label{subsec:EigenvalueCleaningNN}
The second module replaces the Ledoit and Péché RIE with a purely data-driven alternative. It learns a vector-valued function $\boldsymbol{f}\colon \mathbb{R}^{n+1} \to \mathbb{R}^n$ that transforms the empirical eigenvalues $\{\widetilde\lambda_{1},\dots,\widetilde\lambda_{n}\}$ of the lag-transformed returns’ correlation matrix (Eq.~\eqref{eq:lagw}) into the cleaned spectrum $\{\lambda^{\star}_{1},\dots,\lambda^{\star}_{n}\}$. Since the learned filtering depends only on the empirical eigenvalues, it is independent of the associated eigenvectors and hence basis-independent: any orthogonal similarity transform of the input correlation matrix yields the same eigenvalue correction, with the cleaned matrix rotating accordingly.

As shown in Eqs. \eqref{eq:vectvalue}-\eqref{eq:permeq}, the target function $\boldsymbol{f}$ is both non-separable and invariant under any permutation of its inputs.  Applying a fully connected layer directly to the unordered vector of empirical eigenvalues would require handling $O(n!)$ possible input orderings, which is plainly infeasible. Deep Set architectures \cite{zaheer2017deep} and attention-based models \cite{vaswani2017attention} enforce permutation symmetry and thus avoid this factorial blow-up, yet neither fully exploits the structure of the problem. Deep Sets adopt a top-down view: each eigenvalue interacts with a global pooled representation of the spectrum. Although universal approximation theorems guarantee expressiveness \cite{wagstaff2021universal}, modelling fine interactions would require an exploding number of pooling tokens as $n$ increases. Attention mechanisms implement a bottom-up approach that begins by computing pairwise interactions. Capturing higher-order dependencies then requires stacking multiple attention–feed-forward modules, which deepens the network, exacerbates vanishing or exploding gradients \cite{liu2020understanding,dong2021attention}, and poses significant computational challenges as $n$ grows.

To design an architecture that more effectively mitigates the curse of dimensionality, we exploit the exact mathematical mapping provided by the Coulomb-gas formulation of RMT \cite{forrester2010log}, which models sample eigenvalues as one-dimensional charges under three forces: mutual logarithmic repulsion, confinement by the population spectrum, and hard-edge repulsion at zero.  The confinement potential reflects the full population spectrum: when eigenvalues are well separated relative to the effective sample length $\Delta t_{\mathrm{in}}$, it is dominated by the nearest population eigenvalue; otherwise, it arises from a complex superposition of bulk contributions. The mutual repulsion between sample eigenvalues scales as $|\widetilde\lambda_k-\widetilde\lambda_l|^{-1}$ and therefore decays with rank separation, while the hard-edge term enforces nonnegativity but weakens as $\Delta t_{\mathrm{in}}$ approaches $n$. Consequently, the shift of each $\widetilde\lambda_k$ away from its population counterpart is governed primarily by interactions with the sample eigenvalues that lie within its local spectral vicinity, a region whose extent depends on the local spacing around $\widetilde\lambda_k$, while the cumulative influence of all more distant eigenvalues can be approximated as a coarse-grained contribution.
Although the resulting spectral density and eigenvector-overlap dynamics could in principle be derived by solving the coupled Hamiltonian equations associated with this analogy, we instead incorporate only the implied locality as an inspiration for a structural prior in our neural architecture, which should lie, in fact, between the extremes of Deep Sets and attention architectures.

To clarify the last statement, consider that within any subset of sample eigenvalues, the effective order of interaction is not fixed in advance but grows with the local density of the population spectrum. Regions with tightly spaced population eigenvalues induce strong collective forces that demand modelling higher‐order dependencies among neighbouring sample eigenvalues, whereas sparse regions admit accurate approximation with only low‐order interactions. In a bottom‐up framework such as self‐attention, capturing these varying orders requires explicitly stacking as many attention–feed‐forward layers as the maximum local interaction order, since each layer only learns pairwise or limited higher‐order dependencies. As the dimension $n$ increases, a region of given spectral concentration spans more eigenvalues, forcing deeper networks to maintain the same modelling fidelity. Consequently, this scaling renders pure attention architectures impractical in very high dimensions.

A Recurrent Neural Network (RNN) can naturally capture such locality. Viewing the rank index as a pseudo-time coordinate, we process the ordered sequence $\widetilde\lambda_{1}\le\dots\le\widetilde\lambda_{n}$ with a Bidirectional Long Short-Term Memory (BiLSTM) network \cite{gers2000forget,schuster1997bidirectional}, which is a specific architecture of the RNN family. 
An RNN employs a single learnable cell function, so that at each step it takes the current input and the previous hidden state to produce an updated hidden state.  This same cell is applied repeatedly across the ordered sequence, allowing the model to process inputs of arbitrary length with a fixed set of parameters. 

For the bidirectional version, the forward and backward cells learn the update functions
\begin{align}
\boldsymbol{g}^\rightarrow(\boldsymbol{h}^\rightarrow_{k-1},\boldsymbol{m}^\rightarrow_{k-1},\widetilde{\lambda}_k,q) &= \{\boldsymbol{h}^\rightarrow_k,\boldsymbol{m}^\rightarrow_k\}, & \text{with} \quad k=1, \ldots, n,\label{eq:forward}\\
\boldsymbol{g}^\leftarrow(\boldsymbol{h}^\leftarrow_{k+1},\boldsymbol{m}^\leftarrow_{k+1},\widetilde{\lambda}_k,q) &= \{\boldsymbol{h}^\leftarrow_k,\boldsymbol{m}^\leftarrow_k\},&  \text{with} \quad k=n, \ldots, 1,\label{eq:backward}
\end{align}
where $\boldsymbol{h}^\rightleftharpoons_k = \left\{ h^{\rightleftharpoons}_{k,1},\dots,h^{\rightleftharpoons}_{k,\omega} \right\}$ and $\boldsymbol{m}^{\rightleftharpoons}_k= \left\{ m^{\rightleftharpoons}_{k,1},\dots,m^{\rightleftharpoons}_{k,\omega} \right\}$ are the hidden and cell states. Setting the input states to zero at the start of the sequences, Eq.~\eqref{eq:forward} is applied from the smallest to the largest eigenvalue, whereas Eq.~\eqref{eq:backward} proceeds in the opposite direction. 

The forward states summarise all eigenvalues to the left of the current rank, and the backward states summarise those to its right. Their concatenation, therefore, encodes a two-sided interaction whose effective range is governed by LSTM gating rather than by an explicit window. Using rank as the temporal axis is unconventional but consistent with the symmetry of long-range yet decaying interactions. By adopting a BiLSTM, the model no longer requires capacity that grows with $n$, as was necessary for both Deep Sets and attention mechanisms. Instead, the BiLSTM learns to approximate the combination of local neighbor interactions and a distant mean-field term. RMT indicates that this combined interaction converges to a fixed filter as $n$ and $\Delta t_{\mathrm{in}}$ increase \cite{ledoit2011eigenvectors}, so model complexity should remain bounded even in very high dimensions.

After scanning the spectrum, the BiLSTM outputs an $n\times2\omega$ matrix $\boldsymbol{H}$ whose $k$-th row $\boldsymbol{h}_{k}$ concatenates column-wise the forward $\boldsymbol{h}^\rightarrow_{k}$ and backward $\boldsymbol{h}^\leftarrow_{k}$ hidden vectors. A dense layer shared across ranks projects $\boldsymbol{h}_{k}$ onto the positive real line through a softplus activation
\begin{equation}\label{eq:NNeigvals}
\lambda_{k,\text{NN}}^{-1}= \operatorname{softplus}(\mathbf a^{\top}\boldsymbol h_k+b),\qquad k=1,\dots,n,
\end{equation}
with trainable parameters $\mathbf a\in\mathbb R^{2\omega}$ and $b\in\mathbb R$. The softplus retains differentiability at the origin and avoids hard saturation, ensuring stable gradients even for very small eigenvalues. To remove the arbitrary overall scale, we normalise such that $\sum_{k=1}^{n}\lambda_{k,\text{NN}}^{-1}=n$. Portfolio weights computed from a RIE do not change if all the eigenvalues are multiplied by the same constant, so this normalization places every instance on the same scale and allows direct comparison in downstream statistical tests without altering the resulting portfolio weights.

In summary, the module comprises a total of $8\omega^2 + 26\omega + 1=34,433$ learnable parameters ($\omega=64$).  Both LSTM cell contributes $8\omega^2 + 24\omega$ parameters, and the final dense layer adds $2\omega + 1$.  If model compactness is a priority, the LSTM units may be replaced by two Gated Recurrent Units (GRUs).  Standard LSTMs employ separate memory and hidden channels, $\boldsymbol{m}$ and $\boldsymbol{h}$, respectively, to mitigate vanishing and exploding gradients.  The cell state $\boldsymbol{m}$, modulated by input, forget, and output gates, can preserve long-range information almost unchanged over many time steps, while the hidden state $\boldsymbol{h}$ undergoes more aggressive step-wise updates and serves as the immediate output. This arrangement improves numerical stability and captures both global and local spectral structure. That said, this dual‐channel design is not strictly necessary: in our experiments, a simplified GRU, which merges long and short‐term memory into a single $\boldsymbol{h}$ performed comparably well. The GRU reorganizes its gating mechanism so that the hidden state alone can reliably encode long-range dependencies, offering a lighter yet effective alternative.  Substituting LSTM cells with GRUs reduces the total parameter count to $6\omega^2 + 19\omega + 1$.

\subsection{Multilayer Perceptron for Inverse Marginal Volatilities}\label{subsec:DeepNN}
The second branch of the network estimates the inverse marginal volatilities from the sample standard deviation of the transformed return series (Eq.~\eqref{eq:lagw}). i.e.,
\begin{equation}
    \widetilde{\sigma}_{i} = \sqrt{\frac{1}{\Delta t_{\text{in}}}\sum_{t=1}^{\Delta t_{\text{in}}} \widetilde{r}_{t,i}^{2} -\left(\frac{1}{\Delta t_{\text{in}}}\sum_{t=1}^{\Delta t_{\text{in}}} \widetilde{r}_{t,i}\right)^{2}}.
\end{equation}
Each standard deviation is then passed through an MLP network with $3$ hidden layers of $(64, 32, 16)$ neurons and a \texttt{Leaky-ReLU} activation function. The output of the network is a single neuron with a \texttt{softplus} activation function, which guarantees that the output is positive. In total, this module comprises $2,753$ learnable parameters.  It is important to stress that we process each asset’s estimated volatility independently through the same MLP, taking one scalar at a time rather than feeding the entire vector of $n$ volatilities jointly.  This ensures that the network’s parameters remain fixed and do not scale with the number of assets, while still learning a universal transform applicable to every marginal volatility.
The output of this branch is then standardized by dividing it by the average inverse output of the network, i.e.,
\begin{equation}
    \sigma_{i,\text{NN}}^{-1} \mapsto \frac{\sigma_{i,\text{NN}}^{-1}}{\frac{1}{n}\sum_{j=1}^{n} \sigma_{j,\text{NN}}^{-1}}.
\end{equation}
This normalization is irrelevant for the portfolio weights, but it is useful to compare the output of the network across different samples.

 Although a more sophisticated design could make the volatility transformation context-dependent, adapting to correlation structure or different market regimes, we deliberately adopt this minimal architecture.  Our focus in this work is on the eigenvalue-cleaning of the correlation matrix, and hence, we keep the marginal volatility branch as simple as possible.

\section{Training and Evaluation}\label{sec:Training}

\subsection{Selection of the Investable Universe}\label{subsec:dataselec}
We describe here the procedure used to select assets for investment from the universe of US equities that were listed on the NYSE or NASDAQ at any point from 1 January 1990 to 31 December 2024. This universe, therefore, includes American Depositary Receipts (ADRs) and explicitly excludes funds and exchange-traded funds. To avoid data leakage, for each trading day $t$, we rely only on information available up to $t-1$ when determining the eligible stocks. We further exclude any security for which a delisting occurs within the subsequent $\Delta t_{\text{out}} = 5$ trading days. This procedure is safe because, under SEC rules, issuers must provide at least 10 days' written notice to the exchange before filing Form 25 to delist securities \cite{sec12d22}.

We designed a two‐tiered filtering scheme that operates on both the full five-year calibration window ($\Delta t_{\text{in}}=1200$) and a short, recent time window.  The full-history filters enforce homogeneity of trading activity over the entire five years and thus attempt to remove any stock whose price series is fragmented by late initial public offerings, prolonged trading halts, or significant corporate restructurings.  The recent-window filters, by contrast, apply stringent liquidity thresholds over the last five to twenty trading days with the attempt to eliminate any security whose executed auctions fall below the levels compatible with low-impact rebalancing.  This dual approach both addresses the challenges of assembling a broad universe of $n=1000$ stocks with uninterrupted five-year histories and ensures that each selected asset remains sufficiently liquid at the time of investment. That said, the threshold values themselves are inherently arbitrary and have been chosen primarily to maintain a stable universe of $n=1000$ stocks at every point in time; nevertheless, filtering approaches like this reflect standard practice in hedge funds, where inclusion criteria must satisfy stringent and complex rules. In practice, working with a very large, unconstrained dataset can introduce unintended biases and degrade the reliability of both risk estimates and optimized allocations.

Our full history filter enforces a homogeneous trading record across the five-year calibration window by requiring that, within any rolling one-year window, a stock has participated in the closing auction on at least 95\% of trading days. By doing so, we guarantee that at every point in time within the year-long window, the vast majority of price observations are present.

On the recent trading history,  the closing auction must have executed on each day, and on each of those days, the daily traded volume must equal at least 1 \% of shares outstanding and 1 \% of market capitalization. This ensures that the stock is liquid enough to be traded with a low impact on the market. The day before the trade, we check that the number of shares outstanding is greater than 5M, to avoid micro-shares. Finally, on the day before trading, we require that each stock’s price lies between 10 \$ and 2,000 \$. Stocks trading at relatively low price levels incur high per-share transaction costs, whereas those at relatively high price levels often exhibit reduced trading activity, impeding efficient execution.

Since our aim is to design a multivariate methodology to obtain a portfolio of stocks with a minimum variance, we decided to filter out those stocks that univariately exhibit an extremely low variance when compared to the universe of stocks. In the presence of those cases, the minimal variance becomes trivial, and the multivariate nature of the problem is lost. In other words, we are interested in building a portfolio of high-variance stocks, which, when combined, can lead to a low-variance portfolio. To this end, we applied a standard procedure to remove outliers. We removed from the available universe those stocks whose distribution of log standard deviation is below the 1.5 interquartile range (IQR) of the log standard deviation of the universe~\cite{dekking2005modern}. This condition must be satisfied in two time windows of 5 and 20 most recent days.

Finally, to reduce redundancy, we keep only one class of shares for each company, the one with the largest market capitalization on the previous day.
A similar filter is executed for those stocks with an in-sample correlation greater than 0.95 \cite{engle2019large}. The final selection is done by ranking the stocks by their market capitalization and selecting the top $n=1,000$ stocks for every trading day $t$. The final panel, spanning January 1990 to December 2024, contains roughly $2.7$ millions of stock–day pairs and approximately $4,000$ unique stocks.

\subsection{Training Procedure}\label{subsec:TrainingProcedure}
The proposed architecture is designed to be independent of the number of assets $n$. 
This means that the same model can be applied to different subsets of assets without the need for retraining. However, we know from RMT that the filtering should depend on the aspect ratio $q_\textrm{in}:=n/\Delta t_\text{in}$; therefore, during the training phase, the NN should be fed with examples of different $n$ sizes if we aim to obtain a model that generalizes well to larger portfolio sizes.
Accordingly, at each batch we randomly and uniformly sampling the number of assets $n$ from a $[50,350]$ range in every batch of the training.  This resampling also enlarges the effective training set, because the number of possible cross sections grows combinatorially and repeating the exact same asset combination becomes unlikely when the investable universe is large. The number of days $\Delta t_{\text{in}}$ is fixed to $1200$ trading days; in fact, the presence of the module of Sec.~\ref{subsec:ReturnsReweighting} includes $2 \Delta t_{\text{in}}$ trainable variables $(\alpha_t, \beta_t)$, which are associated with specific past lags. In the future version, this could be generalized to a variable number of past lags, but it would require a more complex architecture.

We trained our model on a dataset of US equities spanning January 1990 to December 2024. For each validation date $t$, we define the calibration window as the period from the dataset start through December 31 of the year preceding $t$. In particular, we conduct validation over the interval [2000-01-01, 2024-12-31], resulting in 24 distinct trained NNs.
To construct a robust training procedure, we randomly draw a date $t$ such that the interval $[t-\Delta t_{\mathrm{in}}-1,t+\Delta t_{\mathrm{out}}]$ is fully contained within the calibration window. We then randomly select $n$ assets from the investable universe at $t-1$ (see Section \ref{subsec:dataselec}). The NN is trained on the adjusted close-to-close returns over the input window $[t-\Delta t_{\mathrm{in}},t-1]$ together with the realized OOS covariance matrix $\boldsymbol{\Sigma}_{\mathrm{out}}$, computed from the adjusted close-to-close returns over $[t,\,t+\Delta t_{\mathrm{out}}]$.
A one-day shift between the training inputs and the OOS returns is introduced to avoid data leakage, since the return at $t$ depends on the closing price at the end of day $t$, which is not available intra-day.

According to the training procedure, the probability that NN will be trained on the same dataset is very low for $n$ sufficiently smaller than the investable universe on a given day. Consequently, continuously monitoring validation-set performance for early stopping becomes less critical. However, to further reduce the risk of overfitting, we used a decay factor of the learning rate, which is set to $10^{-4}$ and drops every batch by a factor of $0.99^{1/500}$, and we clip the gradients by norm to $1.0$. The gradient is then optimized with the Adam optimizer with an automatic rescaling of the loss function. 
We trained all our models with $100$ epochs, with $500$ steps per epoch, and a batch size of $32$ samples. To ensure that the convergence is stable, for each of the 24 years, we trained our NN 10 times independently, so finally, we obtained 240 trained NNs.  

All training runs were executed on CPU, without GPU acceleration, and did not pose any computational bottlenecks. On a standard Intel i7 processor, one epoch (500 steps with batch size 32) requires approximately 200 seconds on average. After calibration, inference is comparatively lightweight, computing portfolio weights for a single instance  requires on the order of $10^{-1}$ seconds on the same hardware.

\begin{figure}[htb]
    \centering
    \includegraphics[width=0.49\linewidth]{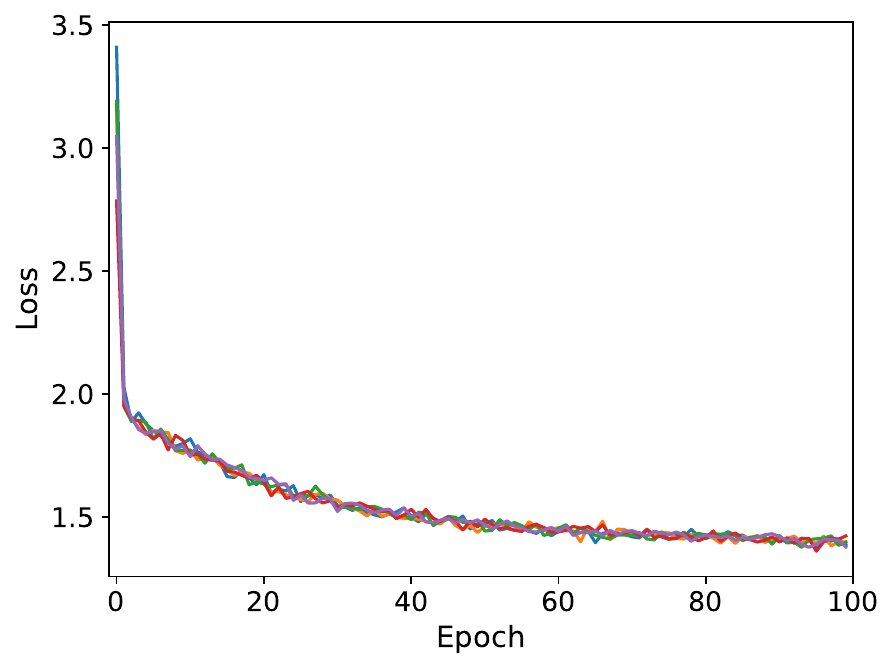}
    \includegraphics[width=0.49\linewidth]{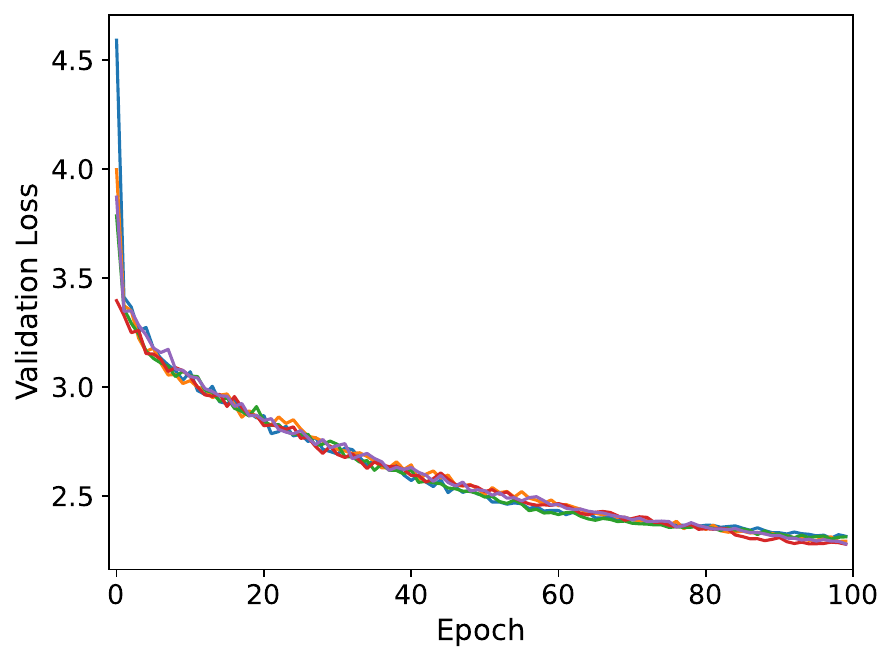}
    
    \caption{Training loss on the left panel, validation loss on the right panel. Different lines refer to independent training runs on the same calibration window. The figure refers to the calibration of the first model up to 31-12-1999. The loss function is defined in Eq.~\eqref{eq:loss}. }
    \label{fig:loss}
\end{figure}

For further evidence of robustness against overfitting, Fig.~\ref{fig:loss} displays the evolution of training and validation losses over successive epochs. The loss trajectories remain consistent across independent runs on the same data interval. Both training and validation losses decrease monotonically. A monotonic decline of the training loss is expected; the validation loss, by contrast, often increases once overfitting sets in. After $60$ epochs, the training loss plateaus, exhibiting an average change of less than $3.5\%$ over the final forty epochs. The validation loss continues to decline at a faster pace, falling by approximately $6\%$ during the same period. These results indicate that terminating training around $100$ epochs would probably incur minimal overfitting risk.

\FloatBarrier

\subsection{Model Interpretability}\label{subsec:ModelInterpretability}
\subsubsection{Lag-Transformed Returns}
\begin{figure}[hbt]
\centering
\includegraphics[width=0.45\columnwidth]{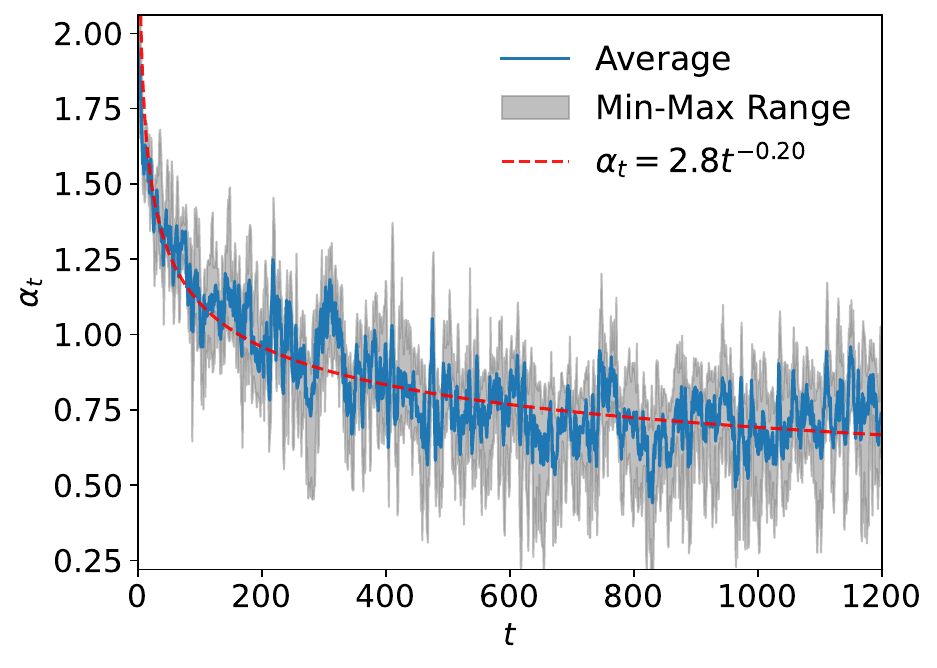}
\includegraphics[width=0.45\columnwidth]{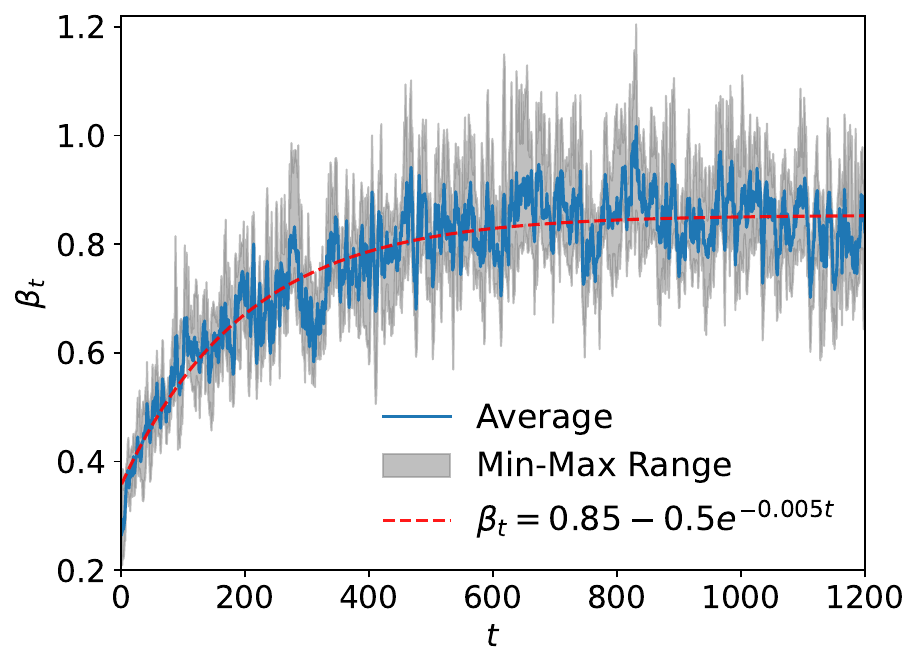}

\includegraphics[width=0.45\columnwidth]{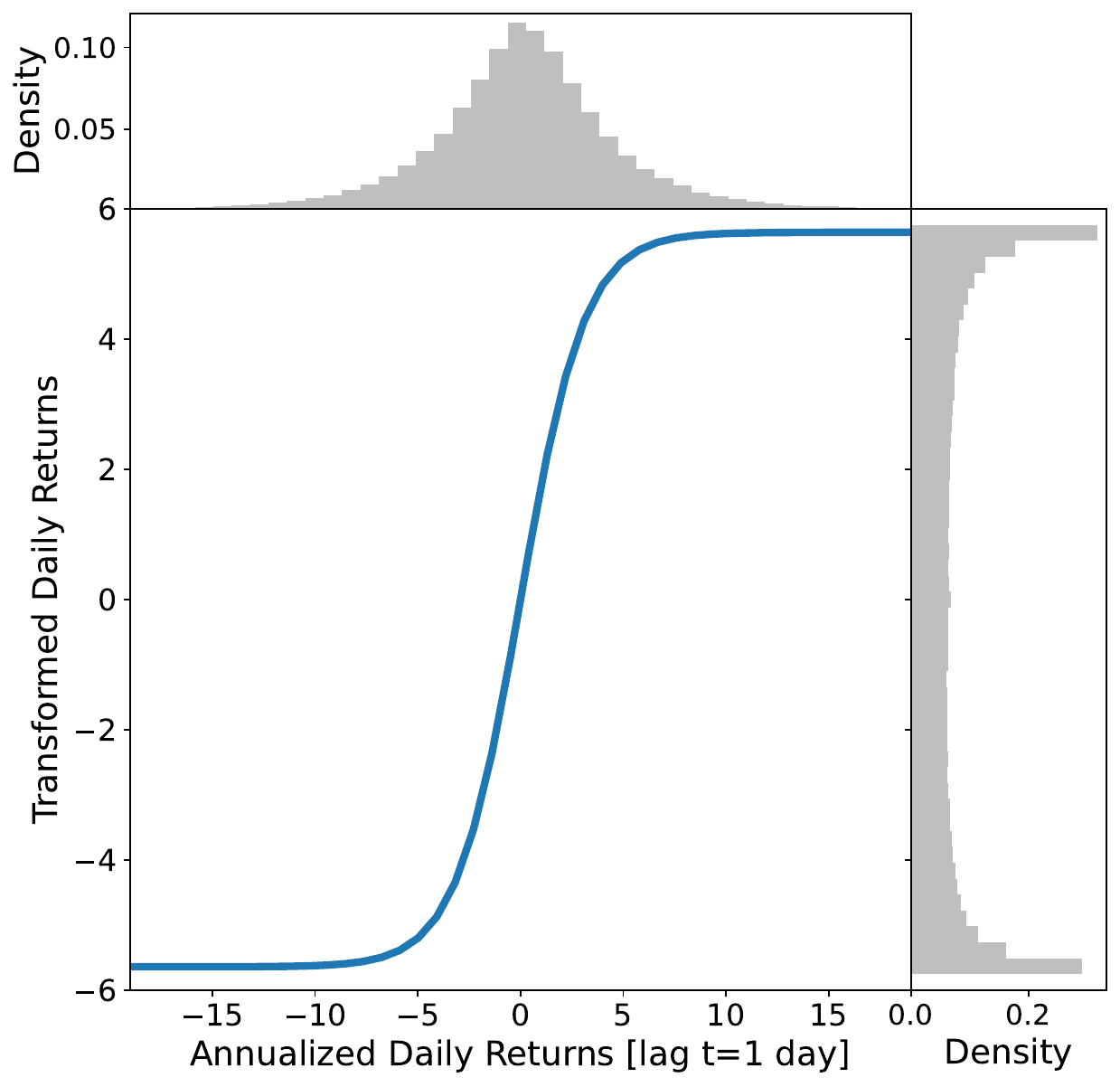}
\includegraphics[width=0.45\columnwidth]{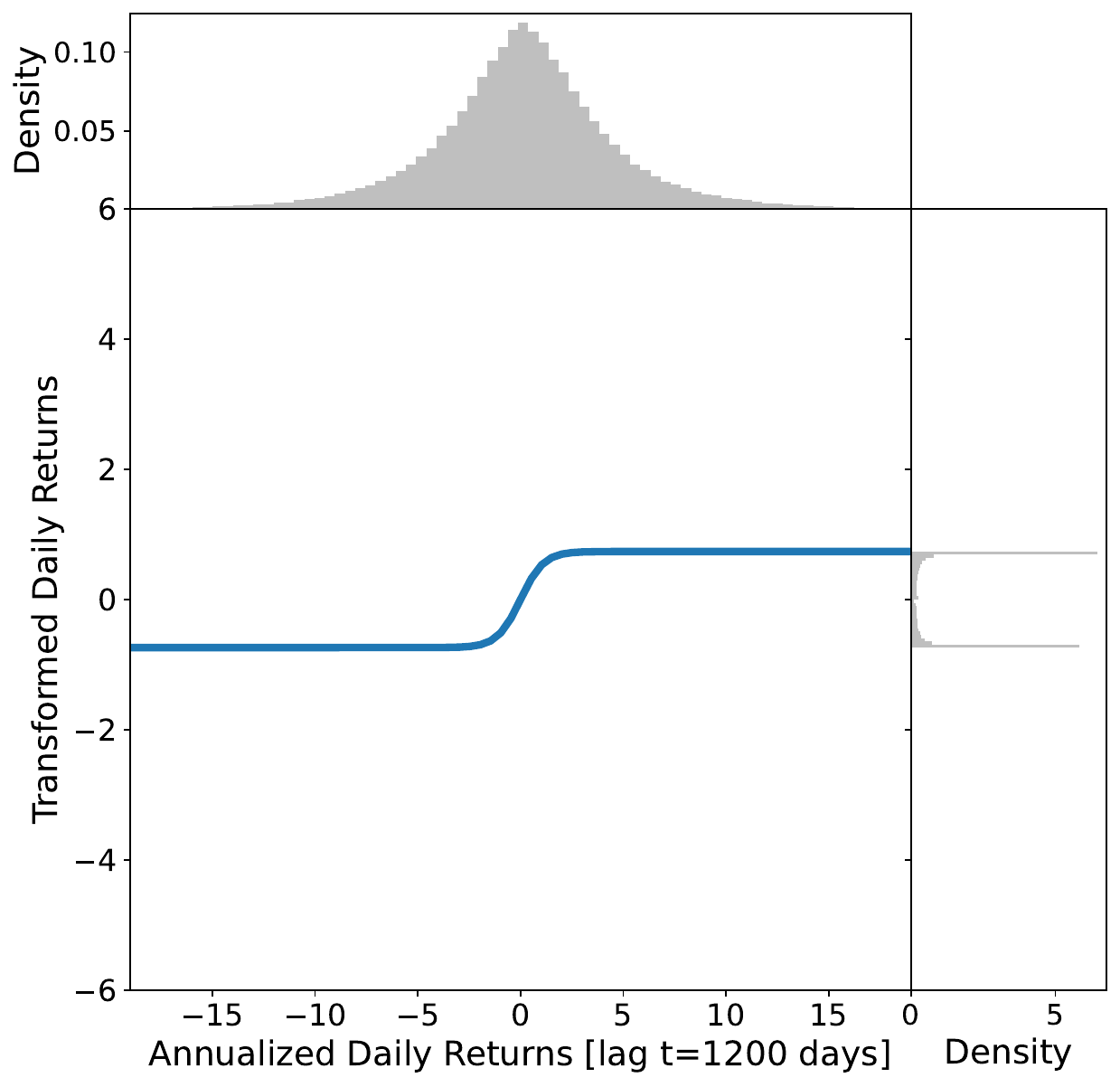}

\caption{The upper plots show the calibrated weighting factors $\alpha_t$ and $\beta_t$ as a function of the lag. The factors shown here are an average over the 240 trained models. Low values of the $x$-axis correspond to the most recent past, while high values correspond to the oldest past. Both figures show an attempt to fit them with a power law for $\alpha_t$ and with a saturating exponential for $\beta_t$ indicated with the dashed red lines. Min-Max ranges highlight the consistency of the results over the 240 independent training. The lower plots show an example of the return transformation of the annualized returns for the last trained model over past examples. The left plot refers to the day before the investment, and the right plot refers to the first day in the dataset. The marginal distribution shows how the empirical distribution is transformed.}
\label{fig:reweighting_factors}
\end{figure}

After the calibration, we can analyze the learned weighting factors $\alpha_t$ and $\beta_t$ of Eq.~\eqref{eq:lagw} as a function of the lag $t$ to understand how the NN is weighting and transforming the past returns. 
Given that we ran $10$ independent training runs for each of the $24$ years of the validations, we obtained $240$ trained models. We can then compute the average of the weighting factors over the trained models, which are shown in Fig.~\ref{fig:reweighting_factors}.

Interestingly, the rescaling factor $\alpha_t$ seems to follow a power law with an exponent of approximately $-0.2$, which in substance means that the most recent past is weighted slightly more than the oldest past. In fact, half of the contribution is given from the $40\%$ most recent past. The emergence of the power law is quite surprising, being the most widely used decay factor based on Exponentially Weighted Moving Average (EWMA) \cite{riskmetrics1996,tan2025estimation}, whereas the power law decay would instead imply a Hyperbolic Weighted Moving Average (HWMA). A model based on HWMA was proposed in Ref.~\cite{kawakatsu2021simple}, but to our knowledge, it is not very popular yet. 

The clipping factor $\beta_t$ shows values within the range $[0.4,1.0]$. Interestingly, the clipping factor $\beta_t$ seems to saturate for larger lags. To have an idea about the range of clipping, for the farthest data points, assuming $\alpha_t=1$, a daily return of $1\%$ is clipped to approximately $0.4\%$ while the same return in the most recent past is clipped to approximately $0.9\%$.

The lower panels of Fig.~\ref{fig:reweighting_factors} present the combined impact of the shrinkage and sigmoid transformations. In the remote past, the transformed values not only contract toward zero but also assume effectively binary states. As a result, beyond approximately one year of lag the series conveys only the sign of daily returns. When the Pearson correlation is applied to a binarized series, it coincides with the Phi coefficient \cite{almog2015mesoscopic,yule1912methods}. In contrast, the most recent returns still exhibit a strong binary character yet develop a pronounced uniform region around the center. This feature reflects the fact that the sigmoid kernel closely tracks the empirical cumulative distribution. By transforming the raw data via the empirical CDF, the Pearson correlation of the transformed series becomes the Spearman coefficient \cite{espana2024kendall}. However, this exact equivalence holds only when each return series is rank-transformed by its own empirical CDF, whereas using a single merged CDF across all assets only approximates the Spearman coefficient and remains inexact due to volatility‐driven distortions across assets.
In summary, as the lag increases, the network’s operative correlation metric shifts continuously from a Spearman‐type measure at short lags toward a Phi‐type measure at long lags.

The above analysis also clarifies an architectural constraint that affects the choice of the IS window length. In the current implementation, the lag-transformation block assigns a distinct pair of parameters $(\alpha_t,\beta_t)$ to each look-back lag $t\in\{1,\dots,\Delta t_{\mathrm{in}}\}$ (Eq.~\eqref{eq:lagw}), so the number of trainable variables in this module scales as $2\Delta t_{\mathrm{in}}$. Consequently, $\Delta t_{\mathrm{in}}$ must be fixed at training time and is kept unchanged at inference. At the same time, the learned shapes in Fig.~\ref{fig:reweighting_factors} suggest a natural route toward flexible window lengths, since $\alpha_t$ and $\beta_t$ appear to follow smooth, low-complexity patterns (approximately a power law and a saturating exponential, respectively), one could replace the per-lag free parameters by a low-dimensional parametric kernel. Such a formulation would decouple the number of learnable parameters from $\Delta t_{\mathrm{in}}$ and would allow a variable look-back horizons within the same architectural family.

\FloatBarrier

\subsubsection{Eigenvalue Sensitivity Analysis}
The counterintuitive result of the AO estimator \cite{bongiorno2023filtering,bongiorno2024covariance} is that a set of fixed eigenvalues can be used to obtain an estimator for the correlation matrix that outperforms all the RMT-based estimators for portfolio optimization. Moreover, such an estimator does not depend on the assets used in the calibration or the exchange, country, or time period. The authors of Ref.~\cite{bongiorno2023filtering} hypothesized that the non-stationarity of the system is so strong that, unless eigenvectors are corrected, very little information can be extracted from the sample eigenvalues. 
In our case, we used an NN trained to attempt extracting specifically the non-stationary relationship between $\boldsymbol{C}_{\text{in}}$ and $\boldsymbol{C}_{\text{out}}$. If the above statement is true, and the NN does not overfit the calibration data, we should expect an output that is independent of the assets in the input. 

\begin{figure}[htb]
\centering
\includegraphics[width=0.45\columnwidth]{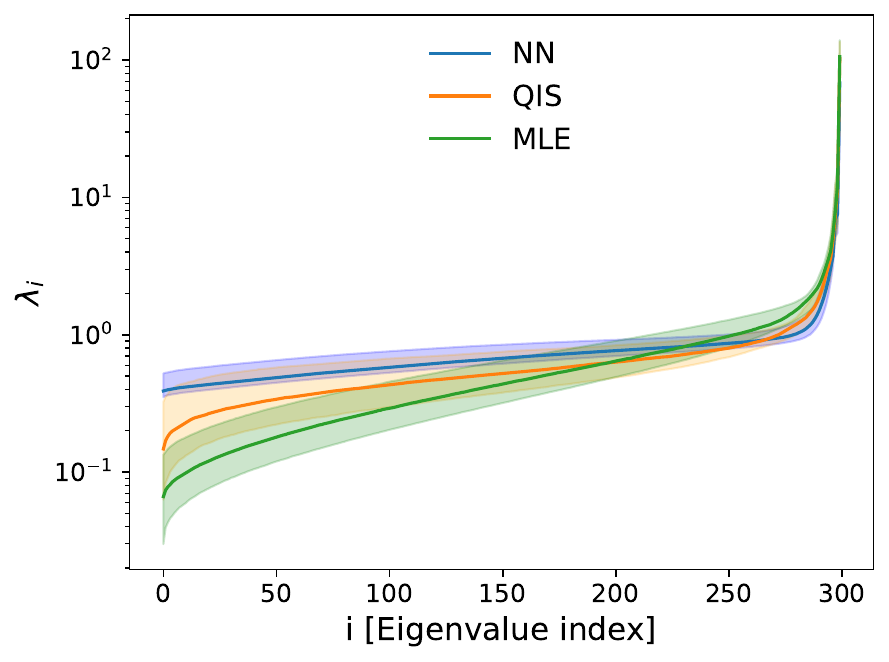}
\includegraphics[width=0.45\columnwidth]{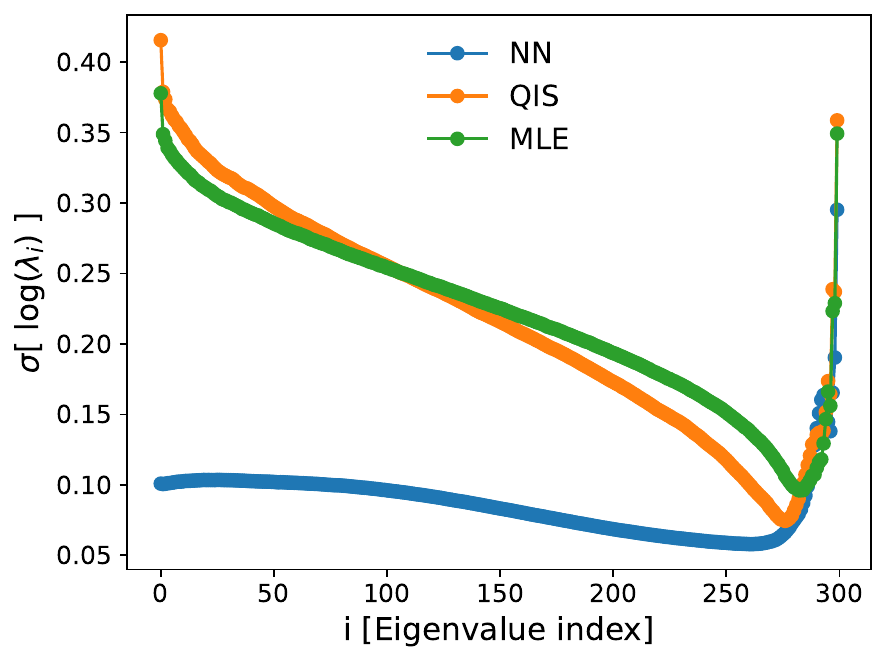}
\caption{Eigenvalue sensitivity analysis. Left: median of the eigenvalues as a function of the rank, the colored bands indicate a $95\%$ percentile coverage. Right: Standard deviation of the log eigenvalues as a function of the rank.  The eigenvalues are computed on 1024 samples from the full calibration dataset.  Abbreviations: NN = Neural Network estimator; MLE = Maximum-Likelihood (sample) Estimator; QIS = Quadratic-Inverse Shrinkage. }
\label{fig:eigenvalue_sensitivity}
\end{figure}

In the left panel of Fig.~\ref{fig:eigenvalue_sensitivity}, we present the eigenvalue spectra for the MLE, the QIS shrinkage, and the NN‐transformed eigenvalues $\boldsymbol{\lambda}_\text{NN}$ after inversion of Eq.~\eqref{eq:NNeigvals}. The range of the largest eigenvalues coincides across all three estimators, whereas the bulk of the spectrum exhibits the most pronounced differences. The MLE assigns the lowest values to the bulk eigenvalues. The QIS shrinkage attenuates this dispersion by smoothing the spectrum. Neither method reproduces the abrupt truncation characteristic of eigenvalue clipping \cite{laloux1999noise}; instead, while both the MLE and QIS yield a bulk spectrum spanning approximately one order of magnitude, the NN transformation compresses the bulk eigenvalues to within roughly half an order of magnitude. This compression causes the NN transformation to resemble more the abrupt truncation of eigenvalue clipping.

The right panel of Fig.~\ref{fig:eigenvalue_sensitivity} shows the standard deviation of the log‐transformed eigenvalues across calibration samples. All estimators exhibit pronounced variability for the largest eigenvalues. In the bulk of the spectrum, the MLE and the QIS continue to display substantial dispersion, whereas the NN‐transformed eigenvalues form a nearly flat plateau. This plateau indicates that the NN mapping renders the bulk eigenvalues effectively insensitive to the input data, akin to the asset‐agnostic behavior of the AO estimator, while still responding to the extreme eigenvalues.

\FloatBarrier

\subsubsection{Marginal Volatility Estimator}

\begin{figure}[hbt]
\centering
\includegraphics[width=0.60\columnwidth]{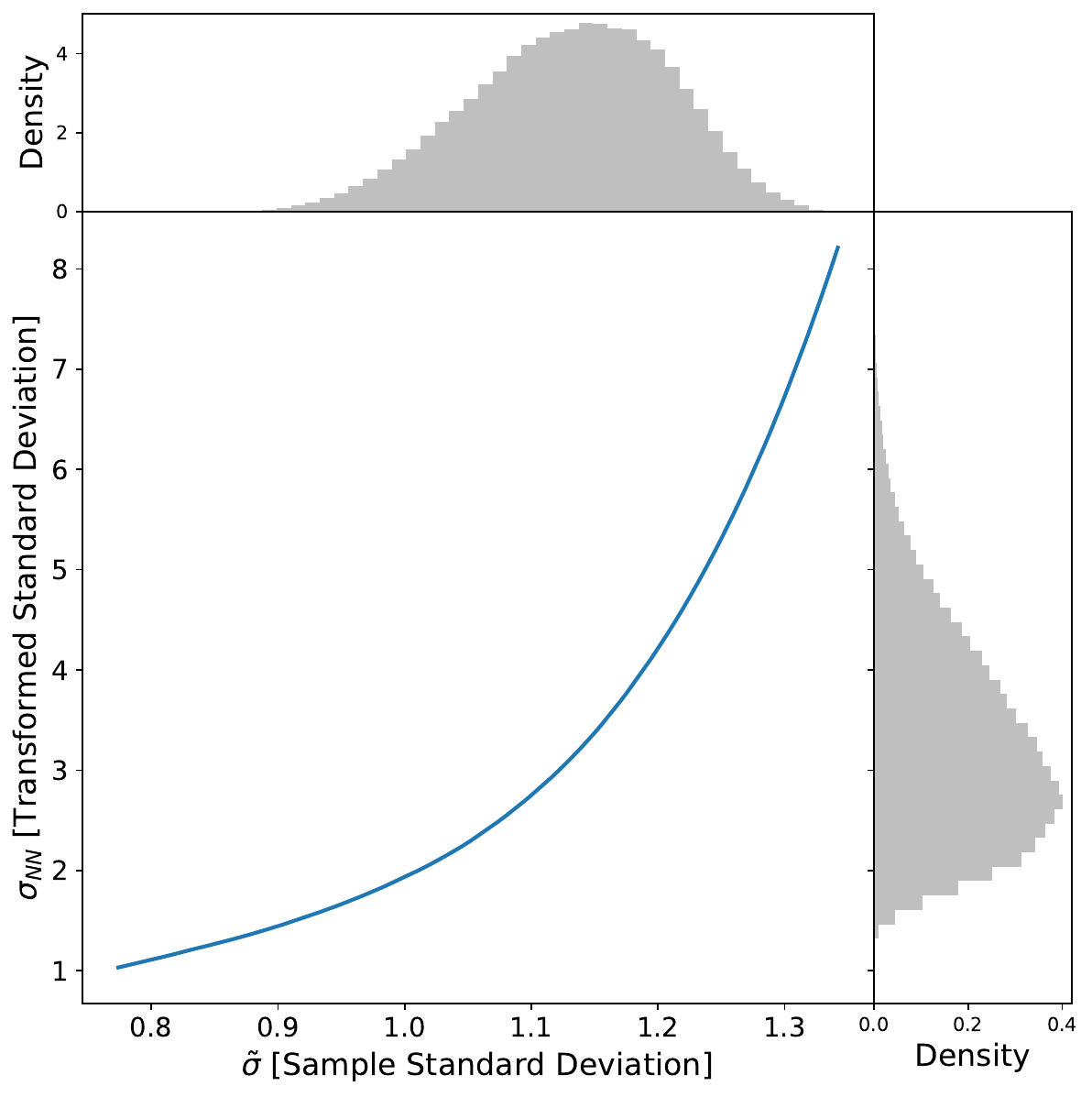}
\caption{The figure shows how the MLP (Model 3) transforms the standard deviation of the lag-transformed returns. The x-axis shows the standard deviation of the lag-transformed returns, while the y-axis shows the output of the NN. The marginal distributions of the input and output are shown on the top and right sides of the figure, respectively.}
\label{fig:inverse_marginal_volatility}
\end{figure}

The second branch of the NN architecture is a deep NN that learns the inverse marginal volatilities. After the training, we can analyze how the NN transforms the standard deviation of the rescaled returns. It is important to notice that this transformation is applied to every standard deviation independently of other stocks and market conditions, and it is also not dependent on the history of the stock itself. Although we believe that this is not the optimal way to learn this variable, the main task of this work is the learning of the eigenvalue cleaning, and this module is set up by purpose as minimally as possible (likewise, the returns lag-transformation module).
In Fig.~\ref{fig:inverse_marginal_volatility}, we show how the NN transforms the standard deviation of the rescaled returns. The x-axis shows the standard deviation in input, and the y-axis shows the output of the NN. The marginal distributions of the input and output are shown on the top and right sides of the figure, respectively. Practically, the NN flattens the left tail of the distribution, which corresponds to low volatility assets, while it amplifies the right tail of the distribution, which corresponds to high volatility assets. This means medium-low volatility assets will have a marginal decrease in volatility, while high volatility assets will have a marginal increase in volatility.

\subsection{Backtesting Procedure}\label{subsec:BacktestingProcedure}
The calibrated network is trained to predict the GMV portfolio by allowing unrestricted short positions.  This design choice yields a closed‐form mapping of the optimal weights via the analytical expression in equation Eq.~\eqref{eq:GMVQP}.  If one imposes a non‐negativity constraint on the weights, the GMV solution must be obtained through a numerical solver.  Although specialized NN layers can implement such constrained quadratic programs with end‐to‐end differentiation \cite{amos2017optnet}, this approach incurs substantially higher computational cost and, in our tests, suffers from vanishing gradients.  On the other side, allowing short sales solely to minimize variance might be undesirable, since it amplifies gross leverage and incurs additional borrowing and transaction costs.

To reconcile analytical tractability with realistic trading constraints, after the uncontrained calibration, we can alter the architecture so that the network directly outputs an estimate of the inverse covariance matrix, $\boldsymbol{\Sigma}_{\mathrm{NN}}^{-1}$ (see Fig. \ref{fig:NN_architecture}).  We invert this matrix and then compute portfolio weights under a long‐only constraint by solving the QP externally.

In Sec.~\ref{sec:frictionless} we evaluate our method against alternative estimators defined in Sec.~\ref{sec:backtest_benchmarks} under idealized, frictionless conditions.  In this experiment, we calibrate the network without trading constraints, i.e., allowing both long and short positions. Then, we test both the raw uncontrained output and the long-only version described above. In Sec.~\ref{sec:realistic} investigates the model’s extrapolation performance.  We train on a small universe of assets and then test on a substantially larger universe, enforcing long‐only positions and employing an operational simulator that incorporates realistic transaction costs. 
\FloatBarrier

\subsubsection{Benchmarking Alternative Estimators}\label{sec:backtest_benchmarks}
We compare two broad classes of portfolio estimators, (i) univariate benchmarks and (ii) multivariate covariance‐cleaning methods, across two complementary experimental designs introduced in Secs.~~\ref{sec:frictionless} and \ref{sec:realistic}. In Sec.~~\ref{sec:frictionless} conduct a bootstrap‐style backtest without transaction costs to enable frictionless, synthetic comparisons over varying universe sizes. In Sec.~\ref{sec:realistic}, we perform a single, realistic simulation on the top‐capitalisation universe of 1,000 stocks, incorporating trading frictions, liquidity constraints, and slippage. Univariate benchmarks rely solely on individual asset variances or market–weight information. Multivariate methods instead apply covariance‐cleaning procedures to the estimated correlation matrix prior to optimization. Although univariate strategies are pervasive in index construction and passive instruments, their ex post performance is predictably inferior; accordingly,  we include them only in our detailed analysis of the realistic backtest of Sec.~\ref{sec:realistic}.

Univariate benchmarks serve as low-information baselines.  The Equal-Risk Budget (ERB) assigns each asset a weight proportional to the reciprocal of its estimated variance, thereby equalising the contribution of each asset to total ex-ante volatility \cite{leote2012demystifying}.  The Market-Capitalization Weighted (MCW) portfolio coincides with the market proxy used in empirical asset-pricing work and therefore absorbs only the cross-sectional information embedded in free-float market values \cite{hsu2004cap}.

Multivariate benchmarks differ only in the way the $n\times n$ correlation matrix $\widehat{\boldsymbol{C}}=\boldsymbol{C}_\bullet$ is obtained.  After the correlation matrix is estimated, we always project on the univariate volatilities $\widehat{\boldsymbol{\Sigma}} = \boldsymbol{D}_\textrm{MLE} \boldsymbol{C}_\bullet \boldsymbol{D}_\textrm{MLE}$ and we apply the QP optimization. Filtering the correlation rather than the covariance is a necessary step to put RIEs in their best performance setup. 
The simplest method is the sample MLE uses the unfiltered sample correlation matrix computed over the look-back window. The Power Mapping (PM), introduced by Guhr and Kälber \cite{guhr2003new}, suppresses estimation noise by raising every off-diagonal element of the de-meaned correlation matrix to a power $\gamma>1$ and re-normalising. 

For the LS, we use the standard implementation of Sklearn \cite{kramer2016scikit} where the parameter is set up according to the original work of Ledoit and Wolf \cite{ledoit2004well}. For QIS, we used a Python implementation released in the GitHub repository \cite{pald22covShrinkage} that we adapted to improve the computational performance.  Finally, AO is calibrated using the universe of stocks as defined in Sec.~\ref{subsec:dataselec} within 1990-2000. Then the eigenvalues are kept constant for the whole validation period, 2000-2024, independently of the stock in the basket. 

We do not aim to provide an exhaustive benchmark of multivariate GARCH/DCC models and their numerous extensions. The DCC literature has produced a wide spectrum of specifications and calibration protocols (see, e.g., Ref.~\cite{sun2025enhancing}), and a systematic comparison across this full class would require a dedicated experimental framework, especially given the well-documented computational overhead and potential convergence instabilities of high-dimensional DCC estimation. More importantly, our contribution targets a different design point: our end-to-end NN (and most benchmarks in this section) produces a single, window-based covariance estimate per rebalancing date and operates within a static eigenvalue-filtering and shrinkage paradigm, rather than learning an explicit state-space evolution for the covariance process. A fully coherent like-for-like comparison would therefore entail extending the NN with modules that ingest and propagate sequences of covariance states, followed by re-benchmarking against the broader family of dynamic models.

However, because DCC-type models remain the canonical econometric benchmark for time-varying correlations since Engle~(2002) \cite{engle2002dynamic}, we include a representative set of state-of-the-art high-dimensional DCC variants in our experiments of Sec.~\ref{sec:realistic}. Specifically, beyond the baseline DCC, we consider shrinkage-regularised DCC implementations in the spirit of Engle, Ledoit, and Wolf~\cite{engle2019large}, where the correlation-targeting matrix is itself denoised via nonlinear eigenvalue shrinkage (QuEST/NLS), and we also test a variant in which this shrinkage step is replaced by quadratic-inverse shrinkage (QIS)~\cite{ledoit2022quadratic}. These specifications are particularly relevant here because they explicitly couple DCC dynamics with rotation-invariant spectral cleaning of the unconditional correlation target. Finally, we include a factor-augmented extension following De Nard, Ledoit, and Wolf~\cite{denard2021factor}, in which an observed Fama-French factor is used to model the dominant market-wide component and DCC dynamics are applied to the residual covariance. This augmentation is designed to stabilise large-universe covariance forecasts by preserving economically meaningful factor directions while still allowing time-variation in the idiosyncratic dependence structure. In related large-universe studies, such hybrid DCC-shrinkage and factor-augmented DCC specifications represent the strongest-performing members of the DCC family, yet they remain challenging in portfolio applications due to instability and turnover effects~\cite{bongiorno2024covariance}. 

\subsubsection{Frictionless Interpolation Backtesting}\label{sec:frictionless}
The first backtesting procedure we used aims to evaluate the performance of the NN within the range of the calibration setup. To avoid confusion, it is important to note that we still mean that the NN is trained in the past and tested in the future, as explained in Sec.~\ref{subsec:TrainingProcedure}.  Instead, we mean that in our training, we explored a range of $n \in [50,350]$ assets, and we are testing the NN on $n=300$ assets that are within the range of the training set. This implies that the NN will interpolate what it has learned during the training phase, and it will not extrapolate to a larger number of assets. Moreover, in this backtesting procedure, we will not include any transaction costs, no slippage, no borrowing costs, or perfect fractional shares. This is the most coherent condition to evaluate if the trained NN was successful in learning the task. In the next section, instead, we will evaluate the NN on a larger number of assets, and we will include realistic conditions. 

To assess the performance in a statistically robust way, we performed a battery of $1,000$ independent backtests. Specifically, every simulation consists of approximately $5$ years of data, starting from a random day $t$ within [2000-01-01, 2019-12-31], and performing $250$ portfolio rebalancing every $5$ trading days. Our model will use the last available calibration obtained the year before the trading day so that no data leakage occurs. The list of assets is randomly sampled from the investable universe (see Sec.~\ref{subsec:dataselec}) on the first day of the simulation, and it is kept fixed if possible for the entire backtest. If some of the assets are excluded from the investable universe during the backtest, they are removed from the portfolio, and another random asset is added to the portfolio so that the number of assets remains constant.
% meno spazio orizzontale fra le colonne
\setlength{\tabcolsep}{4pt}   % da default è ~6pt
% meno “altezza” fra le righe
\renewcommand{\arraystretch}{0.9}  % da default è 1.0

\begin{table}[ht]
\centering

\begin{tabular}{l|c c c c c c c}
\toprule
\textbf{Method} & $\mathcal{L}(\boldsymbol{w}, \boldsymbol{\Sigma}_\mathrm{out})$ & \textbf{Vol.} & \textbf{Return} & \textbf{Sharpe} & $\mathbf{n_{\mathrm{eff}}}$ & \textbf{Turn.} & \textbf{GL} \\
\midrule \midrule
\multicolumn{8}{l}{\textbf{Unconstrained (Long\,--Short)}} \\ \\
NN   & \textbf{0.0149} & \textbf{10.9\%} & \textbf{11.0\%} & \textbf{1.011} & 15.2 & 112.0\% & 3.07 \\
QIS  & 0.0151          & 11.1\%          & 10.5\%          & 0.942          & 10.6 &  44.0\% & 3.81 \\
AO   & 0.0184          & 12.1\%          & \textbf{11.0\%} & 0.907          & \textbf{28.6} &  \textbf{16.0\%} & \textbf{2.13} \\
PM   & 0.0158          & 11.4\%          & 10.2\%          & 0.902          & 10.3 &  53.0\% & 4.00 \\
LS   & 0.0163          & 11.5\%          & 10.3\%          & 0.893          &  7.2 &  71.0\% & 4.73 \\
MLE  & 0.0168          & 11.7\%          & 10.2\%          & 0.868          &  6.5 &  78.0\% & 5.02 \\
\midrule
\multicolumn{8}{l}{\textbf{Constrained (Long\,--Only)}} \\ \\
NN   & \textbf{0.0234} & \textbf{13.5\%} & 10.7\%          & \textbf{0.792} & 16.5 &  47.0\% & 1.00 \\
AO   & 0.0274          & 14.6\%          & \textbf{10.8\%} & 0.740          & \textbf{33.2} &  \textbf{10.0\%} & 1.00 \\
LS   & 0.0257          & 14.2\%          & 10.3\%          & 0.723          & 13.6 &  14.0\% & 1.00 \\
PM   & 0.0259          & 14.3\%          & 10.3\%          & 0.721          & 17.6 &  13.0\% & 1.00 \\
MLE  & 0.0257          & 14.2\%          & 10.2\%          & 0.721          & 13.3 &  14.0\% & 1.00 \\
QIS  & 0.0258          & 14.2\%          & 10.2\%          & 0.719          & 13.5 &  14.0\% & 1.00 \\
\bottomrule
\end{tabular}
\caption{Performance comparison of frictionless portfolio backtests for both unconstrained (long-short) and constrained (long-only) settings. The table reports averages over $1{,}000$ independent simulations of $250$ rebalancings (every 5 trading days) on portfolios of $n=300$ stocks randomly sampled during [2000-01-01,2019-12-31], without transaction costs. Values are sorted by Sharpe ratio, and bold entries indicate the statistically significant top rank (percentile bootstrap, $p<10^{-6}$). Portfolio volatility, mean return, and Sharpe ratio are annualized; $n_{\mathrm{eff}}$ is computed via the inverse Herfindahl index \cite{woerheide1992index}. The loss function $\mathcal{L}(\boldsymbol{w},\boldsymbol{\Sigma}_\mathrm{out})$ is defined in Eq.\eqref{eq:loss} and refers to the short-term variance ($\Delta t_\textrm{out}=5$ days). Turnover is measured excluding price drift, and gross leverage is one by definition under long-only constraints. Abbreviations: NN = Neural Network estimator; MLE = Maximum-Likelihood (sample) Estimator; QIS = Quadratic-Inverse Shrinkage; LS = Linear Shrinkage; PM = Power Mapping; AO = Average Oracle; Vol. = annualized Volatility; Turn. = Turnover; GL = Gross Leverage; $n_{\mathrm{eff}}$ = Effective Number of assets (inverse Herfindahl index).}\label{tab:frictionless_backtests}
\end{table}

In Tab.~\ref{tab:frictionless_backtests}, we report the average performances of the models tested on the same set of assets. The first column displays the short-term portfolio variance, which coincides with the neural network loss function defined in Eq.~\eqref{eq:loss}. As expected, our trained model outperforms all other estimators in the metric that it optimizes.
 The unconstrained (long–short) results show that the NN achieves the highest Sharpe ratio (1.01) with an annualized volatility of $10.9\%$ and a mean return of $11.0\%$.  The QIS estimator attains the second‐best Sharpe ratio ($0.94$) but does so with a notably high gross leverage of $3.81$, compared with the NN leverage of 3.07.  The AO method ranks third in Sharpe ratio ($0.91$) yet operates with the lowest leverage ($2.13$) and minimal turnover ($16.3\%$).
These performance gains for the NN and QIS come at the cost of trading intensity.  The network rebalances more than its entire notional every five trading days (turnover $112.0\%$), and QIS rebalances $44.0\%$ of its portfolio at each rebalance.  Such high turnover and leverage would incur significant financing and borrowing costs not captured in this frictionless experiment. 

Once short-selling is forbidden (bottom of Tab.~\ref{tab:frictionless_backtests}), most of the models perform similarly. This is a known behavior of covariance cleaning for portfolio optimization \cite{pantaleo2011improved,jagannathan2003risk}. The annualized volatility is around $14.2\%$ for all the models, including the MLE covariance estimator, which is the non-cleaned covariance matrix. The AO estimator performs slightly worse, reaching $14.6\%$ annualized volatility. The NN instead reaches $13.5\%$ annualized volatility, which is the lowest among all the models by a substantial margin. In terms of the Sharpe ratio, the NN reaches a $0.79$, which is the highest among all the models, while the AO estimator reaches the second-best Sharpe ratio of $0.74$, while all the other methods range around $0.72$. We want to stress that checking the Sharpe ratio is very important when measuring the performance of a portfolio in a large, uncontrolled setting. In fact, it might be possible that the data cleaning included stocks that must be excluded, for example, stocks that are about to be delisted or merged or stocks that are not liquid enough. In those cases, the asset might have a very low volatility and a very low return. A portfolio that includes those assets will have a very low volatility, but it will also have a very low return, leading to a very low Sharpe ratio. The objective of a GMV portfolio is, in fact, to minimize the volatility of investing in risky assets. 

The other two columns in Tab.~\ref{tab:frictionless_backtests} report the average number of effective assets in the portfolio. This metric is dominated by the AO estimator, which has an average of $33.2$ effective assets, while the others range between $13.3$ and $17.6$, which are less than half of the number of assets in the portfolio. 

Finally, the average turnover reported in the last column measures the fraction of assets that are changed in the portfolio at every rebalancing. The AO estimator has the lowest turnover ($10\%$) while the NN has a turnover of $47\%$, which is the highest among all the models. This is due to the fact that the NN is trained to be reactive and to focus on the most recent past, while the other models are more stable and less sensitive to recent changes. This behavior is a double-edged sword; on one side, having a more reactive portfolio allows one to adapt to the market conditions, but on the other side, it leads to a higher turnover and, therefore, higher transaction costs. For this reason, in the next section, we will evaluate the performance of the NN in a more realistic setting. 
\FloatBarrier

\subsubsection{High Realistic Extrapolation Backtesting}\label{sec:realistic}
The aim of this backtesting procedure is to evaluate the performance of the NN on a larger number of assets with long-only constraints and to include realistic conditions in our simulator. The main concern of the previous backtesting procedure (Sec.~\ref{sec:frictionless}) is that the NN tends to have a high turnover, which can lead to high transaction costs and slippage. 

The simulator described in this section reproduces, on a daily grid, the complete life cycle of an Interactive Brokers (IBKR) cash-and-margin account. All computations are performed in continuous calendar time, but market-sensitive events, price moves, dividends, corporate actions, and interest accruals are aligned to the exchange calendar supplied with the price data. The model is intentionally operational: every quantity that appears in a real IBKR statement (net liquidation value, margin excess, tiered credit and debit interest, commissions, SEC fees, clearing charges, etc.) is reproduced with the same algebra and the same day-count conventions used by the broker. No frictions are left unmodelled; conversely, no speculative optimization (such as VWAP execution, limit orders, or intraday crossing) is assumed so that the simulation yields a lower-bound estimate of attainable performance.

At any date $t$, the model keeps a position vector $\mathbf{s}_t = \left\{s_{t,i}\right\}_{i=1}^n$ expressed in shares and a free cash $c_t$  scalar. Cash dividends are credited automatically and increase $c_t$. If free cash turns negative after trading, the account is deemed in debit. Interest accrues daily on the debit balance according to the IBKR tiered schedule \cite{interactivebrokers_commissions}. Conversely, when $c_t$ is positive, the broker pays credit interest on the portion above successive notional caps according to the corresponding credit-tier spreads; however, since the first \$100{,}000 tranche is unpaid, we do not model that. To determine the daily rate, a 360-day convention is used. IBKR defines its USD reference rate $\phi^{(r)}_t$ as the Fed Funds Effective Rate i.e., the volume-weighted average of interbank transactions processed through the Federal Reserve \cite{interactivebrokers_reference_rate_2025}, so we set  $\phi_t = \phi^{(r)}_t + \delta_v$ where $\delta_v$ is the volume-dependent spread published by IBKR \cite{interactivebrokers_rate}.

At a rebalancing date, the target number of shares is determined from the estimate of Net Liquidation Value $\widehat{\text{NLV}}_t$. To compute the NLV, we need an estimate for the price of the equities at the closing auctions. We use the primary-exchange opening price if available; failing that, it defaults to the fully split-adjusted close of the preceding session. This choice mimics the information set of a manager who submits market orders before the end of the auctions, accepting execution price risk. The NLV pre-execution is then computed as
\begin{equation}\label{eq:NLV_pre}
\widehat{\text{NLV}}'_t = c_{t-1} + \sum_{i=1}^n s_{t-1,i} \, \widehat{p}_{t,i},
\end{equation}
Please note that $\boldsymbol{s}_{t-1}$ is the position vector corrected by eventual splits. The target number of shares is then computed as
\begin{equation}
    \boldsymbol{s}_t = \operatorname{round}\left( \frac{\mathbf{w}_t \, \widehat{\text{NLV}}'_t}{\widehat{\mathbf{p}}_t} \right).
\end{equation}
Finally, the actual $\text{NLV}_t$ post-execution is computed with
\begin{equation}\label{eq:NLV_post}
\text{NLV}_t = c_t + \sum_{i=1}^n s_{t,i} \, p_{t,i},
\end{equation}
where $\boldsymbol{p}_t$ is the actual close price of the day, and $c_t$ is already discounted for the dividend accrued during the day and fees.

The difference between the target number of shares $\boldsymbol{s}_{t}$ and the previous-day position $\boldsymbol{s}_{t-1}$, corrected for eventual splits, is used as an end-of-day market order request $\Delta \boldsymbol{s}_t = \boldsymbol{s}_{t} - \boldsymbol{s}_{t-1}$. The trades are executed with the actual close price of the day $\boldsymbol{p}_t$, without accounting for market impact. The executed volume is tracked to set up the proper tier of commissions and fees, and it is reset at the beginning of each month. Executed trades incur three additive costs. First, commissions follow the IBKR tiered schedule for US equities: a per-share rate of 0.035\% when the month-to-date routed volume is below 300,000 shares and 0.020\% above that threshold, with a floor of \$0.35 per ticket \cite{interactivebrokers_commissions}. Second, exchange, clearing, and regulatory fees are modeled as a flat 0.0145\% FINRA trading activity fee plus 0.05\% exchange and 0.02\% clearing fees, for a total of 0.0845\% of notional executed. Third, ``Section 31 SEC'' fees apply to the sell notional at 1.157 basis points \cite{interactivebrokers_commissions}. All charges are debited instantaneously from free cash.

The simulator returns, for every calendar date, the full path of $\text{NLV}_t$, cash, market values, and margin statistics, thus reflecting the realistic interaction between turnover, leverage due to slippage, and transaction costs. The simulation starts with a cash balance of 1,000,000 \$, which is a reasonable amount for a small hedge fund portfolio, and it is sufficiently low to avoid the need to account for market impact on the most liquid US stocks \cite{salek2023price}. Our portfolio is rebalanced every $5$ trading days, and the investable asset universe is taken from the procedure described in Sec.~\ref{subsec:dataselec}, with the only difference that we select the top $n=1,000$ for marketcap, defined from the previous day's close price. In this case, we are attempting to extrapolate the NN to a larger number of assets, given that the calibration was done on a maximum of $350$ assets. Furthermore, $q:=n/\Delta t_\text{in} \approx 0.83$, which is substantially larger than the training range of $q \in [0.04,0.29]$.

Evaluating the trained model on a larger universe highlights the practical value of the dimension agnostic design. Because the parameters are shared across assets and eigenvalue ranks, a network calibrated on moderate cross sections can be deployed at inference on substantially larger panels without retraining, which reduces the training cost and simplifies model maintenance when the investable universe changes. This extrapolation test also provides evidence that the learned spectral transformation captures a stable, transferable regularization mechanism rather than idiosyncrasies of a specific panel.

\begin{table}[ht]
\centering
%\scriptsize
\begin{tabular}{l|rrrrrllrr}
\toprule
\textbf{Method} & $\boldsymbol{\mathcal{V}_5}$ & \textbf{Sharpe}& \textbf{Sortino} & \textbf{Mean} & \textbf{Vol.}  &\textbf{VaR$_{5\%}$} &\textbf{CVaR$_{5\%}$}& \textbf{Turn.} & $\mathbf{n_{\mathrm{eff}}}$ \\
\midrule \midrule
NN   & \textbf{9.0\%}   & \textbf{1.058} & \textbf{1.268} & 12.6\% & \textbf{11.9\%}  &\textbf{-1.05\%} &\textbf{-1.76\%}& 57.0\% & 46.0 \\
AO   & 9.7\%            & 0.942          & 1.155          & 11.7\% & 12.5\%           &-1.12\% &-1.85\%& 18.0\% & 47.8 \\
PM   & 9.8\%            & 0.870          & 1.077          & 10.8\% & 12.5\%           &-1.13\% &-1.84\%& 23.0\% & 25.7 \\
QIS  & 9.8\%            & 0.848          & 1.053          & 10.6\% & 12.6\%           &-1.10\% &-1.84\%& 23.0\% & 20.1 \\
 DCC& 10\%& 0.930& 1.177& 11.5\%& 12.4\% &-1.1\% &-1.80\%& 54\%&17.0\\
MLE  & 9.8\%            & 0.847          & 1.052          & 10.6\% & 12.5\%           &-1.10\% &-1.84\%& 25.0\% & 18.8 \\
ERB  & 20.2\%           & 0.533          & 0.710          & \textbf{13.0\%} & 24.5\%   &-2.41\% &-3.59\%& \textbf{6.0\%}  & \textbf{580.2} \\
MCW  & 15.7\%           & 0.484          & 0.615          & 9.4\%  & 19.5\%           &-1.88\% &-2.95\%& 7.0\%  & 146.6 \\
 \bottomrule
\end{tabular}
\caption{Performance of the long-only portfolio with $n=1000$ top capitalization stocks, rebalanced every 5 trading days, from January 2000 to December 2024. The simulations use a realistic trading simulator that accounts for transaction costs and slippage. Ratios, returns, and volatility are annualized. The turnover metric is computed without accounting for price drift and serves only as a reference to the model's stability, $n_\text{eff}$ is computed by using the inverse Herfindahl index \cite{woerheide1992index}. The first column indicates the average short-term portfolio variance ($\Delta t_\textrm{out}=5$) consistently with the NN loss function (Eq.~\eqref{eq:loss}). In addition to variance-based metrics, we report tail-risk measures (VaR and CVaR) computed from the realised portfolio return series at the 5\% left tail. Values are sorted by Sharpe ratio,  and bold values indicate the best-ranked model. Abbreviations: NN = Neural Network estimator; MLE = Maximum-Likelihood (sample) Estimator; QIS = Quadratic-Inverse Shrinkage; PM = Power Mapping; AO = Average Oracle; DCC = Dynamic Conditional Correlation; ERB = Equal-Risk Budget; MCW = Market-Capitalization Weighted; VaR$_{5\%}$ = 5\% Value-at-Risk; CVaR$_{5\%}$ = 5\% Conditional Value-at-Risk; Vol. = annualized Volatility; Turn. = Turnover; $n_{\mathrm{eff}}$ = Effective Number of assets (inverse Herfindahl index).}\label{tab:Realistic}
\end{table}

In Tab.~\ref{tab:Realistic}, we can observe the performance metrics of the portfolio across different strategies. Notably, when considering the realized variance over the five-day OOS window (i.e., the first column of the table), our NN model again achieves the lowest variance, a result that is fully consistent with its calibration objective.  This risk reduction is corroborated by the quantile-based tail measures reported in the table, as both  Value-at-Risk (VaR) and Conditional Value-at-Risk (CVaR) are also improved by the NN. Furtheremore, the NN strategy outperforms the others also in terms of Sharpe ratio, Sortino ratio, and reaches the second-best mean return. The second-best Sharpe ratio is achieved by the AO estimator. The DCC benchmark ranks close behind (Sharpe $0.93$) but remains below both NN and AO. All the other methods, including the MLE covariance estimator, have similar performances in terms of Sharpe ratio, Sortino ratio, and volatility. The univariate methods, such as ERB and MCW, perform significantly worse than the multivariate methods, with a Sharpe ratio of $0.678$ and $0.483$, respectively. In fact, for the ERB strategy, which allocates weights by inverting individual variances in a univariate fashion, the diversification indicator $n_\text{eff}$ reaches 580 out of 1,000, reflecting a relatively uniform allocation. As detailed in Sec.~\ref{subsec:dataselec}, stocks with exceptionally low past volatility were excluded, thereby preserving the multivariate nature of the optimization problem.

Turnover is the main implementation friction in this simulator because rebalancing is executed via end-of-day market orders and all broker-level costs are applied. In Tab.~\ref{tab:Realistic}, the NN displays a turnover of $57\%$ per 5-day rebalance, higher than AO ($18\%$) and the shrinkage-based estimators (about $20$--$25\%$), and comparable to DCC ($54\%$). Despite this trading intensity, the NN remains top-ranked in realised variance and tail-risk measures, which suggests that the learned allocations are robust enough for the net performance to absorb the additional costs. If turnover were still a binding operational constraint (e.g., through tighter margin usage under leverage), it can be mitigated without modifying the network by adding an explicit turnover penalty or weight-change constraint in the external rebalancing step.

\begin{figure}[hbt]
\centering
\includegraphics[width=0.45\columnwidth]{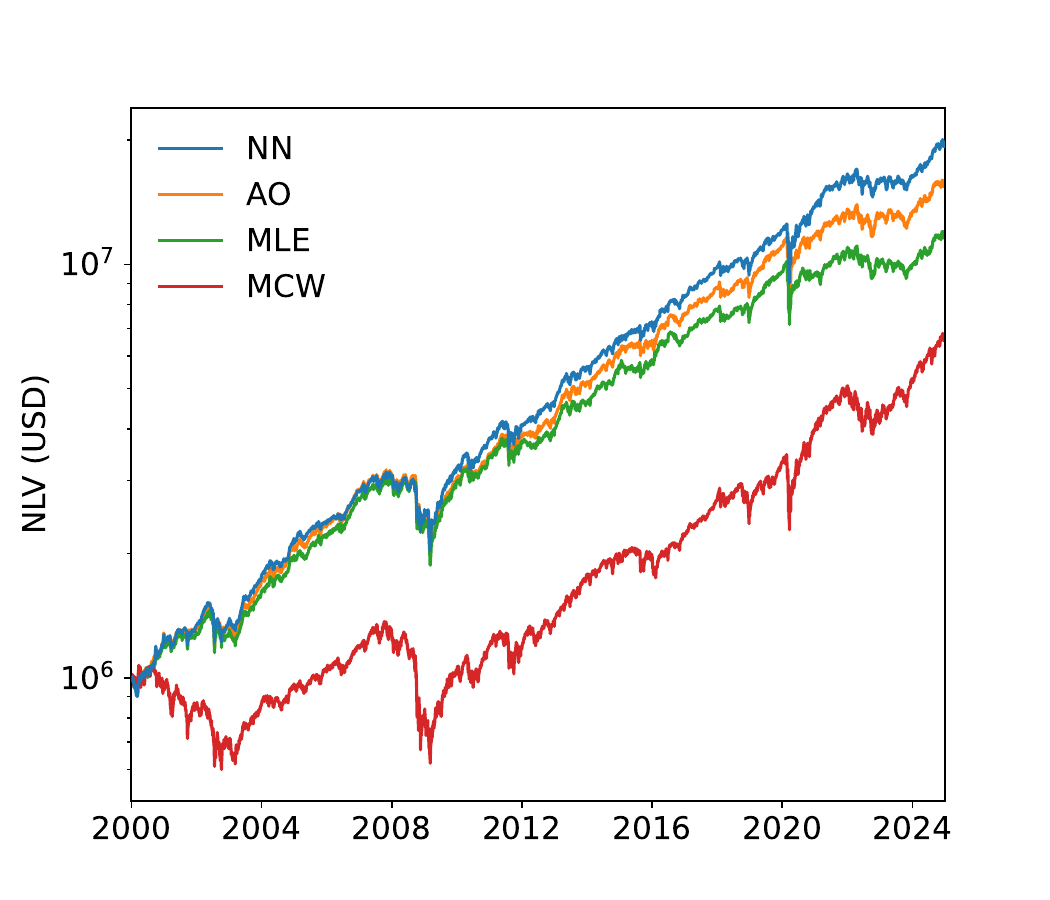}
\includegraphics[width=0.45\columnwidth]{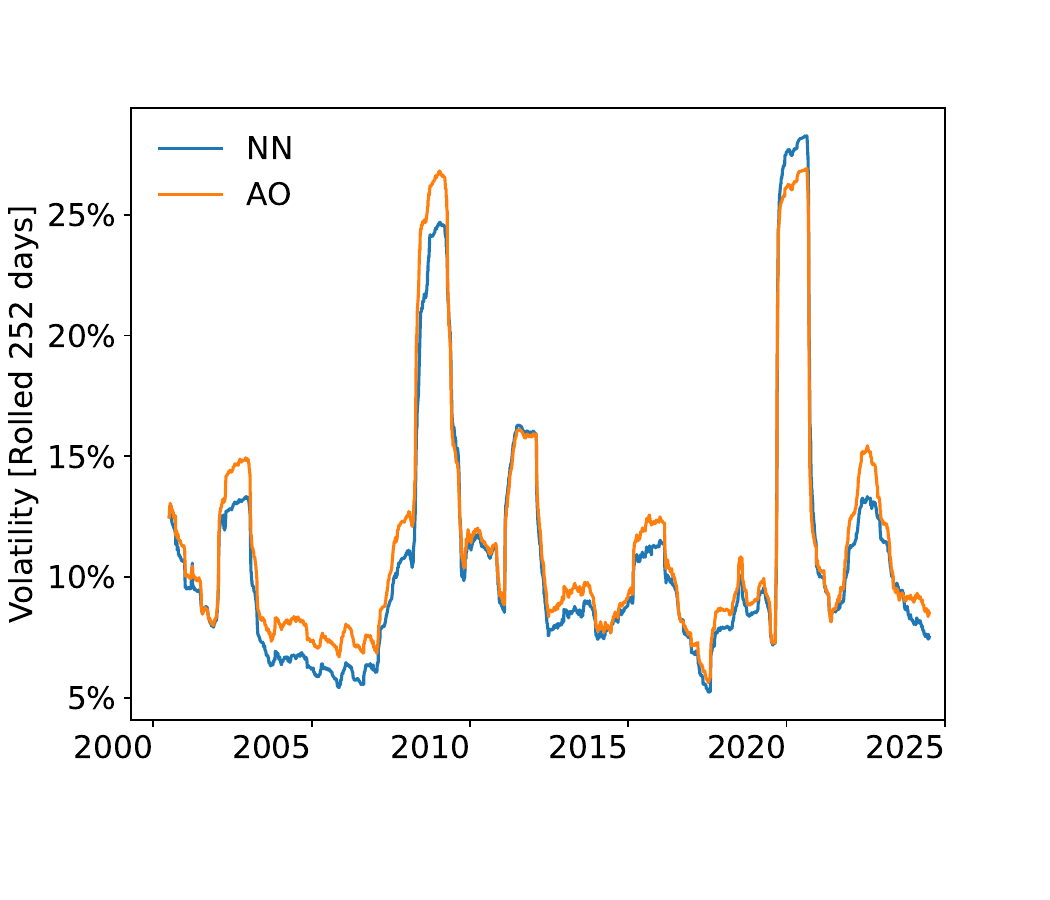}

\includegraphics[width=0.9\columnwidth]{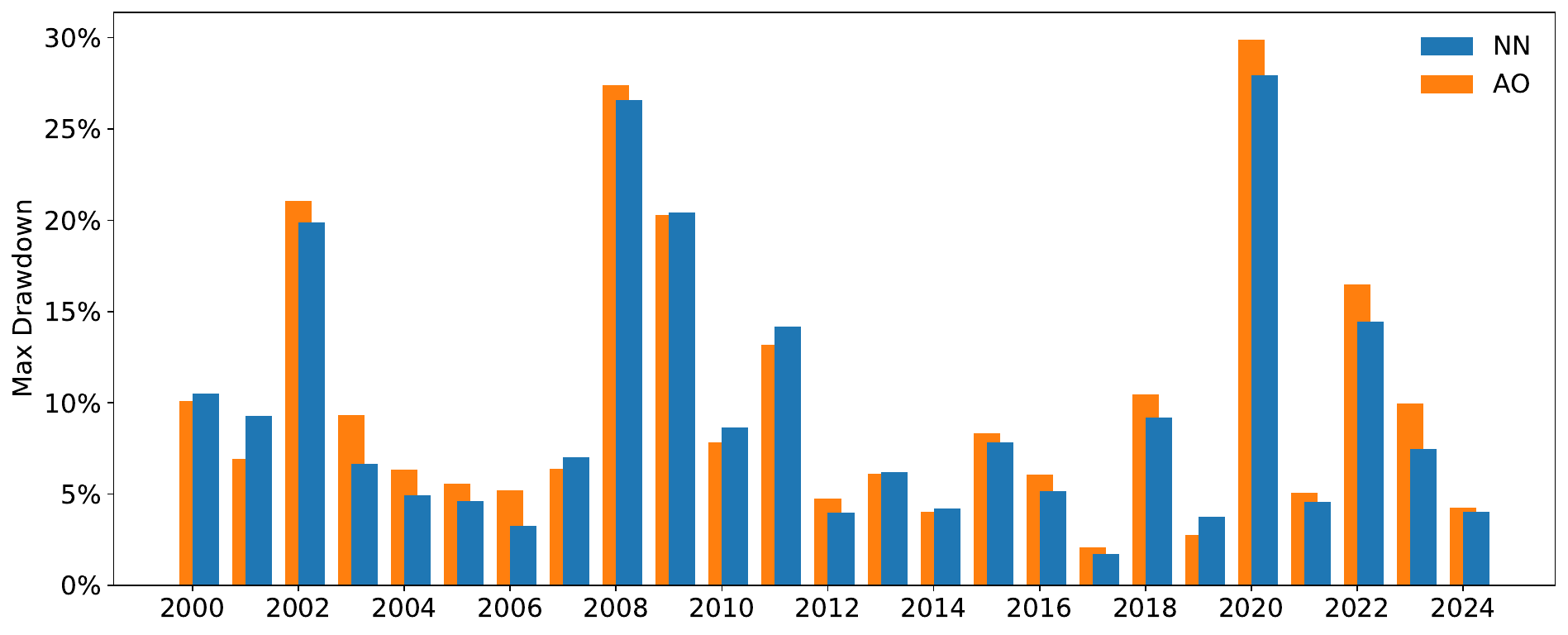}
\caption{Long-only portfolio performances of the top 1,000 most capitalized stocks in the universe backtested with the high realistic simulator. Top left: NLV of the portfolio as a function of time. Top right: Annualized volatility of the portfolio as a function of time over a yearly rolling window. Bottom: MDD of the portfolio by year. Abbreviations: NN = Neural Network estimator; MLE = Maximum-Likelihood (sample) Estimator; AO = Average Oracle; MCW = Market-Capitalization Weighted;  NLV = Net Liquidation Value; MDD = Maximum Drawdown.} \label{fig:NLV_n_1000}
\end{figure}

In the top left panel of Fig.~\ref{fig:NLV_n_1000}, we show the NLV of the portfolio as a function of time. The NLV is computed as in Eq.~\eqref{eq:NLV_post} and it is shown in log scale. From the figure, we can observe that the better performance of the NN in terms of returns is even higher after 2012. This is very important since many NN-based approaches tend to outperform only in the far past, typically before the huge surge of artificial intelligence \cite{fischer2018deep}, but we observe the exact opposite. In the plot, we just focus on four models, the NN, AO, PM, and MLE, which are the ones that are most different from each other according to the performance metrics in Tab.~\ref{tab:Realistic}.  In the top right panel of Fig.~\ref{fig:NLV_n_1000}, we show the annualized volatility of the NN and AO portfolios as a function of time over a yearly rolling window. The NN portfolio has a systematically lower volatility than AO across almost the entire period, with the only exception of 2020, which is the year of the COVID-19 pandemic.  In the bottom panel of Fig.~\ref{fig:NLV_n_1000}, we show the Maximum Drawdown (MDD) of the portfolio by year. Also, this plot confirms that the NN portfolio has a systematically lower or equal MDD than AO, with only a few exceptions.

\FloatBarrier

Finally, we address whether there exists an optimal aspect ratio $q_\textrm{in}:=n/\Delta t_{\mathrm{in}}$ for the scalable model. In our setup, $\Delta t_{\mathrm{in}}$ is fixed, hence varying the universe size $n$ directly sweeps $q_\textrm{in}$.
Tab.~\ref{tab:NN_universe_size} shows that, over the tested range, performance improves monotonically as $n$ increases (equivalently, as $q_\textrm{in}$ increases), with the best results attained at $n=1000$ (corresponding to $q_\textrm{in}\simeq 0.83$). We therefore do not observe an optimal-$q_\textrm{in}$ window in the interior in these experiments. This trend is consistent with the fact that larger universes offer more diversification opportunities, while the main mechanism that would otherwise degrade GMV performance at large $q_\textrm{in}$ is the growth of sampling noise, which our denoising procedure precisely mitigates.

\begin{table}[t]
\centering
%\scriptsize
\begin{tabular}{r|rrrrrllrr}
\toprule
$\mathbf{n}$ & $\boldsymbol{\mathcal{V}_5}$ & \textbf{Sharpe} & \textbf{Sortino} & \textbf{Mean} & \textbf{Vol.} & \textbf{VaR$_{5\%}$} & \textbf{CVaR$_{5\%}$} & \textbf{Turn.} & $\mathbf{n_{\mathrm{eff}}/n}$ \\
\midrule \midrule
21   & 12.6\% & 0.513 & 0.669 & 8.0\%  & 15.5\% & -1.44\% & -2.31\% & 18.3\% & 56.6\% \\
55   & 11.4\% & 0.575 & 0.752 & 8.1\%  & 14.2\% & -1.29\% & -2.07\% & 26.2\% & 19.4\% \\
144  & 10.6\% & 0.680 & 0.847 & 9.1\%  & 13.4\% & -1.20\% & -1.99\% & 29.2\% & 12.1\% \\
379  & 9.9\%  & 0.921 & 1.160 & 11.6\% & 12.6\% & -1.15\% & -1.85\% & 32.4\% & 5.8\% \\
1000 & 9.0\%  & 1.058 & 1.268 & 12.6\% & 11.9\% & -1.05\% & -1.76\% & 57.1\% & 4.6\% \\
\bottomrule
\end{tabular}
\caption{Performance of the NN long-only portfolio as a function of the investment universe size $n$ (top-capitalization stocks), rebalanced every 5 trading days, from January 2000 to December 2024, using a realistic trading simulator that accounts for transaction costs and slippage. Ratios, returns, and volatility are annualized. The turnover metric is computed without accounting for price drift and serves only as a reference to the model's stability. The last column reports the diversification ratio $n_{\mathrm{eff}}/n$ (in percent), with $n_{\mathrm{eff}}$ computed via the inverse Herfindahl index \cite{woerheide1992index}. The first column reports the average short-term portfolio variance at $\Delta t_\mathrm{out}=5$, consistently with the NN loss function (Eq.~\eqref{eq:loss}). Abbreviations:  VaR$_{5\%}$ = 5\% Value-at-Risk; CVaR$_{5\%}$ = 5\% Conditional Value-at-Risk; Vol. = annualized Volatility; Turn. = Turnover; $n_{\mathrm{eff}}$ = Effective Number of assets (inverse Herfindahl index).}
\label{tab:NN_universe_size}
\end{table}

\section{Discussions}
\label{sec:Discussion}
Recent studies discussed in this work illustrate the potential of neural networks in portfolio optimization. They capture nonlinear dynamics, enforce positive definiteness, combine convolutional and recurrent layers, or apply stagewise screening and weighting. They nevertheless share some limitations. They typically operate on low-dimensional or static universes that reduce generality. They employ heterogeneous benchmarks that often omit the GMV portfolio. They may leak future information through survivor-bias filters or delisting exclusions. Finally, the stability of these methods under varying cross-sectional dimensions remains under-investigated, leaving it unclear whether they truly maintain robust performances.

We aim to address these gaps by embedding each analytical step of the GMV solution into a modular neural architecture that remains agnostic to the number of assets. Our network learns the inverse covariance matrix end-to-end by minimizing future portfolio variance. It is calibrated on a few hundred equities and then applied without retraining to one thousand US stocks. The investable universe is selected each day using history-based and liquidity filters that prevent data leakage and exclude assets with anomalously low variance. In backtests with a high-fidelity simulator from January 2000 to December 2024 under realistic trading conditions, our model delivers lower realized volatility, smaller maximum drawdowns, and higher Sharpe ratios than state-of-the-art methods.  Notably, its end-to-end long–short calibration does not pose a serious limitation. In fact, when long-only constraints are imposed on its learned covariance matrix via an external optimizer, its performance advantage over competing estimators remains intact.

The interpretability of our approach follows from its three learnable modules. The lag-transformation block discovers a hyperbolic decay of past returns and applies soft clipping so that recent observations yield a Spearman-type correlation while older returns collapse to a binary sign.  The eigenvalue-cleaning block employs a bidirectional LSTM to perform optimal eigenvalue cleaning. The outcome of this filtering is an approximately input-agnostic shrinkage of the eigenvalue bulk, which is less sensitive than Random Matrix Theory predictions, while eigenvalue outliers are dynamically adjusted by the network. The marginal-volatility block uses a simple feed-forward network that flattens the lower tail of the volatility distribution and inflates the upper tail. These modules reveal how temporal lags, risk modes, and individual volatilities shape the portfolio, thereby providing theoretical insight into the model’s behavior. This interpretability also ensures transparency, aligning with industry standards that discourage opaque black-box solutions.

Despite the empirical improvements documented above, our current architecture remains constrained by the same structural assumptions that underlie RIEs. In particular, the eigenvalue-cleaning module acts only on the empirical spectrum and preserves the empirical eigenvectors, so the method cannot explicitly exploit, nor denoise, the time-varying information carried by eigenvector dynamics (e.g., evolving sector structure or changing factor exposures). As a consequence, the model should not be interpreted as a fully dynamic covariance forecaster or as a mechanism that detects regime shifts in a principled sense. Rather, it provides a data-driven regularisation of finite-sample noise within a static spectral framework, and it can only react to non-stationarity indirectly through the learned lag reweighting and clipping applied to the fixed-length input window. A second practical limitation is that the lag-transformation block allocates separate parameters to each look-back lag, which fixes calibration window length at training time and prevents a genuinely adaptive horizon without architectural changes.

These limitations motivate several concrete extensions. First, one can move beyond rotational invariance by replacing the purely spectrum-based filter with a context-dependent cleaning that is explicitly conditioned on market state variables or on temporal features, thereby enabling genuinely time-varying covariance representations. Second, denoising can be extended to eigenvectors, for instance by incorporating clustering-based structure that modifies both eigenvalues and eigenvectors, as suggested by the empirical success of hierarchical filters. Third, the modules could be coupled so that lag-transformation, eigenvalue filtering, and volatility scaling interact dynamically rather than operating as largely separate components. Finally, we plan to generalize the framework beyond GMV by integrating return estimation and/or differentiable constrained optimization layers, so that the training objective can directly reflect practical constraints (e.g., long-only mandates or turnover control) instead of relying on external post-processing.

% \section*{Author Contributions}
%  C.B.~conceived the study, designed the NN architecture, developed the code implementation, conducted the numerical simulations, and drafted the manuscript. E.M.~contributed to the code development, carried out the exploratory simulations, data analysis, prepared the initial draft of the manuscript, and participated in the revision. R.N.M.~reviewed the methodology and results, participated in discussions to refine the analysis, and contributed to the manuscript revisions.

%Acknowledgments
%\section*{Acknowledgments}
%This work was performed using HPC resources from the ``Mésocentre'' computing center of CentraleSupélec and École Normale Supérieure Paris-Saclay supported by CNRS and Région Île-de-France (\url{http://mesocentre.centralesupelec.fr/}).
%E.M.~acknowledges a fellowship  funded  by PNRR  for the PhD  DOT1608375  in Sistemi complessi per le scienze fisiche, socio-economiche e della vita of Catania University.

%Bibliography
\bibliographystyle{unsrt}  
\bibliography{references}

\end{document}